\documentclass[conference]{IEEEtran}
\IEEEoverridecommandlockouts
\usepackage{cite}
\usepackage{amsmath,amssymb,amsfonts}
\usepackage{algorithmic}
\usepackage{graphicx}
\usepackage{textcomp}
\usepackage{xcolor}
\def\BibTeX{{\rm B\kern-.05em{\sc i\kern-.025em b}\kern-.08em
    T\kern-.1667em\lower.7ex\hbox{E}\kern-.125emX}}

\DeclareMathOperator*{\argmin}{arg\,min}
\usepackage{mathtools}

\usepackage{steinmetz}
\usepackage{authblk}
\newcommand{\minus}{\scalebox{0.75}[1.0]{$-$}}
\usepackage[ruled,norelsize]{algorithm2e}
\makeatletter
\newcommand{\removelatexerror}{\let\@latex@error\@gobble}
\makeatother

\usepackage{blindtext}

\usepackage[skip=2pt,font=scriptsize]{caption}
\usepackage[skip=2pt,font=scriptsize]{subcaption}

\usepackage{scalerel}
\newcommand\wh[1]{\hstretch{2}{\hat{\hstretch{.5}{#1}}}}

\begin{document}

\title{Rate-Splitting Multiple Access for Communications and Jamming in Multi-Antenna Multi-Carrier Cognitive Radio Systems}
\author{Onur~Dizdar,~\IEEEmembership{Member,~IEEE,}
	and~Bruno~Clerckx,~\IEEEmembership{Fellow,~IEEE} 
	\thanks{The authors are with the Department of Electrical and Electronics Engineering, Imperial College London, London, UK.
		(e-mail: \mbox{\{o.dizdar,b.clerckx\}@imperial.ac.uk}.)\\This work was supported by the Engineering and Physical Sciences Research Council of the UK (EPSRC) Grant number EP/S026657/1, and the UK MOD University Defence Research Collaboration (UDRC) in Signal Processing.}
}

\maketitle
\begin{abstract}
With the increasing number of wireless communication systems and the demand for bandwidth, the wireless medium has become a congested and contested environment. 
Operating under such an environment brings several challenges, especially for military communication systems, which need to guarantee reliable communication while avoiding interfering with other friendly or neutral systems and denying the enemy systems of service. 
In this work, we investigate a novel application of Rate-Splitting Multiple Access (RSMA) for joint communications and jamming with a Multi-Carrier (MC) waveform in a multi-antenna Cognitive Radio (CR) system. 
RSMA is a robust multiple access scheme for downlink multi-antenna wireless networks. RSMA relies on multi-antenna Rate-Splitting (RS) strategy at the transmitter and Successive Interference Cancellation (SIC) at the receivers. 
By employing RSMA at the secondary transmitter, our aim is to simultaneously communicate with Secondary Users (SUs) and jam Adversarial Users (AUs) to disrupt their communications while limiting the interference to Primary Users (PUs) in a setting where all users perform broadband communications by MC waveforms in their respective networks.
We consider the practical setting of imperfect CSI at Transmitter (CSIT) for the SUs and PUs, and statistical CSIT for AUs.
We formulate a problem to obtain optimal precoders which maximize the mutual information under interference and jamming power constraints. 
We propose an Alternating Optimization-Alternating Direction Method of Multipliers (AO-ADMM) based algorithm for solving the resulting non-convex problem. 
We perform an analysis based on Karush-Kuhn-Tucker (KKT) conditions to determine the optimal jamming and interference power thresholds that guarantee the feasibility of problem and propose a practical algorithm to calculate the interference power threshold.  
By simulation results, we demonstrate that RSMA achieves a higher sum-rate performance than Space Division Multiple Access (SDMA) and Non-Orthogonal Multiple Access (NOMA).
\end{abstract}

\section{Introduction}
The current and emerging wireless communication systems require broadband communications to meet the data requirements for the advancing applications, both in commercial and military standards. 
As a consequence of the increasing number of systems and applications, the Radio-Frequency (RF) spectrum has become a congested and contested environment.
This leads to an increasing demand for the valuable frequency spectrum and even results in different systems having to contest for or operate in the same frequency band. 

In such a congested and contested RF environment, efficient use and sharing of spectrum is of high importance, especially for military communications with strict reliability and robustness requirements. 
A major challenge for a military network is to operate reliably without interfering with the communications of friendly or neutral networks using the same spectrum, such as the commercial systems 4G and 5G. 
Another critical challenge for military networks arises in the context of Electronic Warfare (EW), which is to prevent the communication of the enemy users, or Adversarial Users (AUs), operating in the same spectrum.

The problem of efficient spectrum sharing calls for interference management capabilities as in Cognitive Radio (CR) networks \cite{mitola_1999}. CR networks enable simultaneous operation of a network of \textit{Secondary Users} (SUs) in the resources originally allocated to the existing \textit{Primary Users} (PUs).
In CR networks, the SUs are responsible to maintain the interference power measured at the PUs under a threshold, or so called interference temperature limit \cite{haykin_2005}, while performing communications in the same spectrum. Under such a constrained environment, the SUs have two conflicting targets of maximizing their throughput and limiting the interference at the PUs. However, SUs can achieve significant throughput by exploiting the resources, such as spatial dimensions, efficiently \cite{zhang_2008}.

On the other hand, denying the \textit{Adversarial Users} (AUs) of their broadband service requires smart and efficient jamming methods. 
In contrast to CR, jamming methods benefiting from Artificial Noise (AN) signals aim to create as much interference as possible at target units in order to degrade their signal quality. 
Among the AN-based jamming methods for Multi-Carrier (MC) broadband waveforms, pilot subcarrier jamming is accepted to be one of the most destructive ones. Pilot jamming aims to disrupt the channel estimation procedure of the target user to prevent error-free detection and decoding of its intended messages \cite{Shahriar_2015, Shahriar_2016, clancy_2011, miller_2012, patel_2004, han_2008}. 

\begin{figure}[t!]
	\centerline{\includegraphics[width=3.2in,height=3.2in,keepaspectratio]{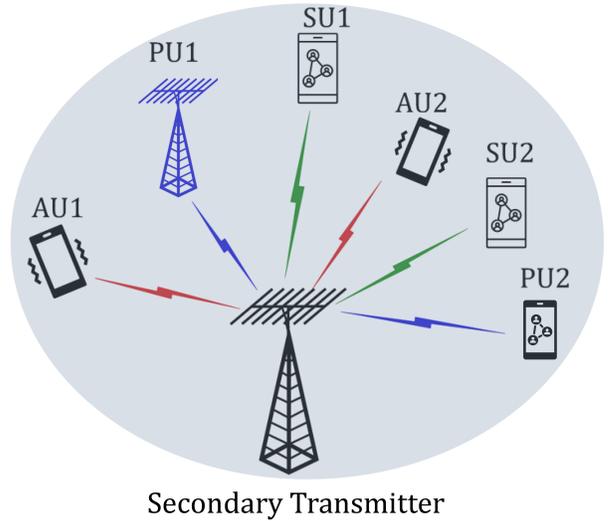}}
	\vspace{0.2cm}
	\caption{System model - we consider a multi-antenna transmitter in the secondary network communicating with SU1 and SU2. The gray area depicts the coverage area of the transmitter and the units in the gray area can receive signal from the transmitter. The units AU1, AU2, PU1, and PU2 communicate in their respective networks with other units outside the coverage area of the transmitter.  The transmitter aims to communicate with SU1 and SU2 while simultaneously disrupting the communications of AU1 and AU2. While doing so, the transmitter aims to limit its interference to PU1 and PU2 in order to avoid disrupting their communications.}
	\label{fig:system}
\end{figure}

The abovementioned operational requirements in an RF-congested environment, as depicted in Fig.~\ref{fig:system}, create three conflicting goals for a communications transmitter: maximizing the throughput in the network while performing jamming to deny the AUs of their service and limiting the interference to PUs to avoid disrupting their communications at the same time.
In this work, we consider Rate-Splitting Multiple Access (RSMA) for multi-antenna multiple-access communications to design an MC system that achieves all the three goals efficiently.
RSMA is a multiple access scheme based on the concept of Rate-Splitting (RS) and linear precoding for multi-antenna multi-user communications. RSMA splits user messages into common and private parts, and encodes the common parts into one or several common streams while encoding the private parts into separate streams. The streams are precoded using the available (perfect or imperfect) Channel State Information at the Transmitter (CSIT), superposed and transmitted via the Multi-Input Multi-Output (MIMO) or Multi-Input Single-Output (MISO) channel \cite{clerckx_2016}. All the receivers then decode the common stream(s), perform Successive Interference Cancellation (SIC) and then decode their respective private streams. Each receiver reconstructs its original message from the part of its message embedded in the common stream(s) and its intended private stream.

RSMA manages multi-user interference by allowing the interference to be partially decoded and partially treated as noise. RSMA has been shown to embrace and outperform existing multiple access schemes, i.e., Space Division Multiple Access (SDMA), Non-Orthogonal Multiple Access (NOMA), Orthogonal Multiple Access (OMA) and multicasting. The sum-rate performance of
RSMA has been demonstrated to be robust and to surpass the performance of SDMA and
NOMA under perfect and imperfect CSIT in numerous works 
\cite{clerckx_2016,clerckx_2019,Joudeh_2016,mao_2018,mao_2019_2}. The performance gain of RSMA has also been studied in single carrier CR systems for Simultaneous Wireless Information and Power Transfer (SWIPT) with perfect CSIT \cite{Acosta_2020}.

In this work, we aim to identify the performance benefits of RSMA for joint communications and jamming in a CR network using MC waveforms with imperfect and statistical CSIT. More specifically, we consider a transmission scheme where the secondary transmitter in a CR network employs RSMA to perform simultaneous communications, jamming and interference management. We obtain optimal precoders that maximize the mutual information for communications between the secondary transmitter and SUs while limiting the interference to PUs and performing jamming on pilot subcarriers of the AUs simultaneously.
We consider the practical and realistic scenario of imperfect CSIT due to the existence of channel estimation for the SUs and PUs, and statistical CSIT for the AUs, since obtaining an accurate channel estimate for the AUs is generally not feasible \cite{karlsson_2017}.

We propose an Alternating Optimization-Alternating Direction Method of Multipliers (AO-ADMM) based algorithm to solve the resulting non-convex problem. 
By means of an analysis based on the Karush-Kuhn-Tucker (KKT) conditions, we show that the thresholds for the interference and jamming power constraints should be set carefully to obtain a non-empty domain for the optimization problem, and determine the thresholds and conditions to guarantee the feasibility of our problem.  
Furthermore, we propose a practical interference power threshold calculation algorithm for given jamming threshold parameters and channel characteristics of the AUs and PUs, which guarantees a non-empty domain for the formulated problem. 

The contributions of the paper can be listed as follows:
\begin{enumerate}
	\item RSMA is considered as the transmission scheme for the secondary transmitter in a multi-antenna multi-carrier CR network to perform simultaneous communications with SUs and jamming of AUs. 
	This is the first paper that considers the application of RSMA in joint communication and jamming.
	Our proposed system model considers the practical case of imperfect CSIT for the SUs and PUs, and statistical CSIT for the AUs. A mutual information maximization problem is formulated under the strict assumptions of imperfect and statistical CSIT with arbitrary numbers of transmit antennas, single-antenna SUs, single-antenna AUs and multi-antenna PUs.  
	\item We perform a feasibility analysis based on Karush-Kuhn-Tucker (KKT) conditions to study the threshold values for the conflicting jamming and interference power constraints in order to obtain a non-empty domain for the formulated optimization problem.
	\item We solve the formulated non-convex problem by an AO-ADMM based algorithm, which consists of an inner and an outer iteration loop. The inner iteration loops are performed by the ADMM algorithm, which solves the rate maximization problem by Mean Square Error (MSE) transformations and makes a projection over the problem domain iteratively. The outer loop is performed by the AO algorithm, which calculates Minimum MSE (MMSE) equalizers for the rate-MSE transformations.
	 We also provide a convergence analysis for the proposed algorithm.
	\item We propose a practical threshold selection algorithm for the interference power constraint under given statistical CSIT for AUs, imperfect CSIT for PUs and jamming power threshold. The interference power threshold values calculated by the proposed algorithm are shown to guarantee the feasibility of the formulated optimization problem. We also show that the proposed threshold detection algorithm returns the optimal threshold values given by the feasibility analysis when the necessary conditions are satisfied and returns a non-trivial threshold otherwise.  
	\item We perform simulations to demonstrate the sum-rate performance of RSMA with the proposed algorithms and compare them with that of SDMA and NOMA as enabling technologies at the secondary transmitter. We use realistic 3GPP frequency-selective channel models obtained from Quadriga to investigate the  performance of the multi-carrier system in a realistic setup. We show that RSMA achieves improved sum-rate performance compared to SDMA and NOMA due its ability to manage interference efficiently and robustness to CSIT imperfections. We also demonstrate the change in the performance of PUs and AUs with and without the proposed system by means of uncoded BER simulations.
\end{enumerate}

The rest of the paper is organized is as follows. Section~\ref{sec:system} gives the system model. We formulate the mutual information maximization problem in Section~\ref{sec:rsma}. 
Section~\ref{sec:proposed} gives the descriptions of the proposed algorithms to solve the formulated non-convex problem.
We give a feasibility analysis for threshold selection and propose an algorithm for calculating the interference power threshold in Section~\ref{sec:threshold}. Section~\ref{sec:simulation} gives the simulation results on the performance of RSMA, SDMA and NOMA with the optimized precoders. Section~\ref{sec:conclusion} concludes the paper. 

\textit{Notation:} Vectors are denoted by bold lowercase letters and matrices are denoted by bold uppercase letters. 
The operations $|.|$ and $||.||$ denote the absolute value of a scalar and $l_{2}$-norm of a vector, respectively, unless stated otherwise. 
$\mathbf{a}^{H}$ denotes the Hermitian transpose of a vector $\mathbf{a}$. $\mathcal{CN}(0,\sigma^{2})$ denotes the Circularly Symmetric Complex Gaussian distribution with zero mean and variance $\sigma^{2}$. $\mathbf{I}_{n}$ denotes the $n$-by-$n$ identity matrix. The operator $\mathrm{tr}(.)$ denotes the trace operation.
The operator $\mathrm{Diag}(\mathrm{X}_{1}, \ldots, \mathrm{X}_{K})$ builds a matrix $\mathrm{X}$ by placing the matrices $\mathrm{X}_{1}$, $\ldots$, $\mathrm{X}_{K}$ diagonally and setting all other elements to zero. The operator $\mathrm{vec}(\mathbf{X})$ vectorizes the matrix $\mathbf{X}$ into a column vector by concatenating its columns.

\section{System Model}
\label{sec:system}
We consider the system model in Fig.~\ref{fig:system}, which consists of a secondary network operating in an RF-congested environment, where the time and frequency resources allocated to the secondary network are also used by several other adversarial and neutral/friendly {\sl primary} networks in the environment. A typical scenario for such a system model consists of a military network for the secondary network, commercial systems ({\sl e.g.}, 5G) as the primary networks and enemy military networks as the adversarial ones. The units in the primary network are referred to as PUs and the ones in the adversarial network are referred to as AUs. We assume that there is no time or frequency resource sharing among the primary, secondary and adversarial networks. 
	
We consider a secondary transmitter in the secondary network with $N_{t}$ transmit antennas serving $K$ single-antenna SUs indexed by $\mathcal{K}=\left\lbrace 1,2,\ldots,K\right\rbrace$. The transmitter performs jamming simultaneously on $L$ single-antenna AUs in its coverage area, which are indexed by $\mathcal{L}=\left\lbrace 1,2,\ldots,L\right\rbrace$.   
The $M$ PUs in the coverage area of the transmitter have multi-antennas, with $N_{r,m}$ denoting the number of antennas at PU-$m$, $ m \in \mathcal{M}=\left\lbrace 1,2,\ldots,M\right\rbrace $.
The secondary transmitter employs an MC waveform to communicate with the SUs, while the AUs and PUs also use MC waveforms to communicate in their corresponding separate networks. We define the set of subcarrier indexes in the signal band as $\mathcal{S}=\left\lbrace 1,2,\ldots,N\right\rbrace $, and the set of pilot subcarriers of AUs-$l$ in the same signal band as $\mathcal{S}_{p,l} \subset \mathcal{S}$, $\forall l \in \mathcal{L}$.  

We consider 1-layer RSMA \cite{mao_2018} at the secondary transmitter to perform multiple-access communications in such a setup. 
RSMA relies on splitting the user messages at the transmitter side. The message intended for SU-$k$ on subcarrier-$n$, $W_{k,n}$, is split into common and private parts, i.e., $W_{c,k,n}$ and $W_{p,k,n}$, $\forall k\in \mathcal{K}$, $n\in \mathcal{S}$. The common parts of the messages of all users are combined into the common message $W_{c,n}$. 
The common message $W_{c,n}$ and the private messages are independently encoded into streams $s^{f}_{c,n}$ and $s^{f}_{k,n}$, respectively. 
Jamming is performed on subcarrier-$n$ of AU-$l$ using the AN signal $s^{e}_{l,n}$, $\forall l \in \mathcal{L}$ and $\forall n \in \mathcal{S}_{p,l}$.
We assume that each subcarrier is assigned a separate precoder. The MC transmit signal at the secondary transmitter using RSMA is written as 
\begin{align}
	x_{n}=\mathbf{p}_{c,n}s^{f}_{c,n}+\sum_{k=1}^{K}\mathbf{p}_{k,n}s^{f}_{k,n}+\sum_{l=1}^{L}\mathbf{f}_{l,n}s^{e}_{l,n}, \  n\in \mathcal{S}.  \nonumber
\end{align}
The vectors $\mathbf{p}_{c,n} \in\mathbb{C}^{N_t}$ and $\mathbf{p}_{k,n} \in\mathbb{C}^{N_t}$ are the linear precoders applied to the common stream and the private stream of SU-$k$ on subcarrier $n$, $\forall k \in \mathcal{K}$ and $\forall n \in \mathcal{S}$. 
The precoder $\mathbf{f}_{l,n}$ is used to transmit AN to AU-$l$, $\forall l \in \mathcal{L}$. The communications signals $s^{f}_{c,n}$ and $s^{f}_{k,n}$ and jamming signals $s^{e}_{l,n}$ are chosen independently from a Gaussian alphabet for theoretical analysis. We also assume that the streams have unit power, so that \mbox{$\mathbb{E}\left\lbrace \mathbf{s}_{n}\mathbf{s}_{n}^{H}\right\rbrace =\mathbf{I}_{K+L+1}$}, where \mbox{$\mathbf{s}_{n}=[s^{f}_{c,n}, s^{f}_{1,n}, \ldots, s^{f}_{K,n}, s^{e}_{1,n}, \ldots, s^{e}_{L,n}]$}. An average transmit power constraint is set as \mbox{$\sum_{n=1}^{N}\mathrm{tr}(\mathbf{P}_{n}\mathbf{P}_{n}^{H})+\mathrm{tr}(\mathbf{F}_{n}\mathbf{F}_{n}^{H})$ $\leq \bar{P}_{t}$}, where \mbox{$\mathbf{P}_{n}=\left[\mathbf{p}_{c,n} \mathbf{p}_{1,n},\ldots,\mathbf{p}_{K,n}\right] $} and \mbox{$\mathbf{F}_{n}=\left[\mathbf{f}_{1,n},\ldots,\mathbf{f}_{L,n}\right] $}. 
Note that instead of assigning separate precoders to each subcarrier, one can assign precoders to each subband, {\sl i.e.,} a group of subcarriers. In this case, $\mathcal{S}$ becomes the index set for the subbands and the problem translates focusing power in specific subbands instead of subcarriers.

The signal received by SU-$k$ on subcarrier-$n$ is 
\begin{align}
	y_{k,n}&=\mathbf{h}_{k,n}^{H}\mathbf{x}_{n}+z_{k,n}, \quad k \in \mathcal{K}, \  n\in \mathcal{S},
	\label{eqn:receive_signal}	
\end{align}
where \mbox{$\mathbf{h}_{k,n} \in \mathbb{C}^{N_{t}}$} is the channel vector of SU-$k$ on subcarrier-$n$ and \mbox{$z_{k,n} \sim \mathcal{CN}(0,N_{0})$} is the Additive White Gaussian Noise (AWGN) component. Similarly, the signals received by the AU-$l$ and PU-$m$ on subcarrier-$n$ are written as 
\begin{align}
	r_{l,n}&=\mathbf{g}_{l,n}^{H}\mathbf{x}_{n}+\nu_{l,n}, \quad  l \in \mathcal{L}, \  n\in \mathcal{S},
	\label{eqn:receive_signal_adversarial}	\\
	t_{l,n}&=\mathbf{M}_{m,n}^{H}\mathbf{x}_{n}+\eta_{m,n}, \quad m \in \mathcal{M}, \  n\in \mathcal{S},
	\label{eqn:receive_signal_primary}
\end{align}
where \mbox{$\mathbf{g}_{l,n} \in \mathbb{C}^{N_{t}}$} is the channel vector of AU-$l$ on subcarrier-$n$, \mbox{$\mathbf{M}_{m,n} \in \mathbb{C}^{N_{t}\times N_{r,m}}$} is the channel matrix of subcarrier-$n$ of PU-$m$ and \mbox{$\nu_{l,n} \sim \mathcal{CN}(0,1)$} and \mbox{$\eta_{m,n} \sim \mathcal{CN}(0,1)$} are the AWGN components.

At the receiver side, detection of the messages is carried out using Successive Interference Cancellation (SIC). The common stream is detected first to obtain the common message estimate $\hat{W}_{c,n}$ by treating the private streams as noise. The common stream is then reconstructed using $\hat{W}_{c,n}$ and subtracted from the received signal. The remaining signal is used to detect the private messages $\hat{W}_{p,k,n}$. Finally, the estimated message for SU-$k$, $\hat{W}_{k,n}$, is obtained by combining $\hat{W}_{c,k,n}$ and $\hat{W}_{p,k,n}$. 
We write the Signal-to-Interference-plus-Noise Ratio (SINR) expressions for the common and private streams at SU-$k$ as
\begin{align}
	\gamma_{c,k,n}\hspace{-0.1cm}=\hspace{-0.1cm}\frac{|\mathbf{h}_{k,n}^{H}\mathbf{p}_{c,n}|^{2}}{N_{0}+Z_{c,k,n}+J_{k,n}}, \quad
	\gamma_{k,n}\hspace{-0.1cm}=\hspace{-0.1cm}\frac{|\mathbf{h}_{k,n}^{H}\mathbf{p}_{k,n}|^{2}}{N_{0}+Z_{k,n}+J_{k,n}}, \nonumber
\end{align} 
\hspace{-0.05cm}with $Z_{c,k,n}=\sum_{i \in \mathcal{K}}|\mathbf{h}_{k,n}^{H}\mathbf{p}_{i,n}|^{2}$, $J_{k}=\sum_{j \in \mathcal{L}}|\mathbf{h}_{k,n}^{H}\mathbf{f}_{j,n}|^{2}$ and $Z_{k,n}=\sum_{i \in \mathcal{K}, i \neq k}|\mathbf{h}_{k,n}^{H}\mathbf{p}_{i,n}|^{2}$.

In this work, we consider the notion of jamming in the context of denial of service for the AUs. Our aim is to efficiently focus power on the AUs to disrupt the correct detection and decoding of their intended data transmissions from other users in their corresponding network. Our performance criterion is the focused power on pilot subcarrier-$n$ of an AU-$l$, $n \in \mathcal{S}_{p,l}$ and $l \in\mathcal{L}$, expressed as 
\begin{align}
	\Lambda_{l,n}=|\mathbf{g}_{l,n}^{H}\mathbf{p}_{c,n}|^{2}+\sum_{k \in \mathcal{K}}|\mathbf{g}_{l,n}^{H}\mathbf{p}_{k,n}|^{2}+\sum_{l^{\prime} \in \mathcal{L}}|\mathbf{g}_{l,n}^{H}\mathbf{f}_{l^{\prime},n}|^{2}.\nonumber
\end{align}
We assume that the transmitter has synchronisation with the AU transmissions \cite{karlsson_2014, karlsson_2017, miller_2012} and a perfect knowledge of $\mathcal{S}_{p,l}$ \cite{Shahriar_2015, Shahriar_2016, clancy_2011, miller_2012, patel_2004, han_2008}.  

We consider the practical case where the transmitter does not have access to perfect Channel State Information (CSI). The channel model of SU-$k$ is expressed as
\begin{align}
	\mathbf{h}_{k,n}=\sqrt{1\minus\sigma_{ie}^{2}}\widehat{\mathbf{h}}_{k,n}+\sigma_{ie}\widetilde{\mathbf{h}}_{k,n},
\end{align}
where $\widehat{\mathbf{h}}_{k,n}$ is the estimate of the channel on subcarrier-$n$ at the transmitter and $\widetilde{\mathbf{h}}_{k,n}$ is the channel estimation error with i.i.d. complex Gaussian elements of unit variance. The entries of $\widehat{\mathbf{h}}_{k,n}$ and $\widetilde{\mathbf{h}}_{k,n}$ are independent.
Similarly, the channel model for the PU-$m$ is expressed as 
\begin{align}
	\mathbf{M}_{m,n}=\sqrt{1\minus\sigma_{pe}^{2}}\wh{\mathbf{M}}_{m,n}+\sigma_{pe}\widetilde{\mathbf{M}}_{m,n},
\end{align}
where the elements of $\wh{\mathbf{M}}_{m,n}$ and $\wh{\mathbf{M}}_{m,n}$ are i.i.d. complex Gaussian elements of unit variance. We assume that an initial CSIT acquisition phase is performed in the secondary network by means of pilot or CSI feedback transmission from the SUs to the secondary transmitter. The pilot transmission can be performed either by means of pilot subcarriers of an MC block, which also carries data from the SUs to the secondary transmitter on the remaining subcarriers; or a dedicated MC block which carries pilots on all subcarriers.
The covariance matrix of the channel of AU-$l$ on subcarrier-$n$ is expressed as
$\mathbf{R}_{l,n}=\mathbb{E}\left\lbrace\mathbf{g}_{l,n}\mathbf{g}_{l,n}^{H}\right\rbrace$. We assume that the channel is fixed during the transmission of an MC waveform block. We also assume perfect CSI at the receivers.

\section{Problem Formulation}
\label{sec:rsma}
In this section, we give a problem formulation to obtain the optimal precoders for the system model in Section~\ref{sec:system}. 
Our objective is to maximize the ergodic mutual information under imperfect CSIT for SUs while focusing a certain amount of jamming power on the pilot subcarriers of the AUs and limiting the interference to PUs. 
The receiver employs carrier non-cooperative processing of the MC waveform. Such approach considers an independent processing of each subcarrier of the waveform at the receiver. Carrier non-cooperative approach is less general in terms of processing of the received signal than the cooperative counterpart, which can cope with intermodulation terms among the subcarriers. However, non-cooperative processing is more suitable for practical scenarios due to its lower complexity \cite{Palomar_2003}.

\subsection{Ergodic Mutual Information with MC Waveforms}
\label{sec:ergodic_mutual}
We define the matrices $\mathbf{H}_{k}=\mathrm{Diag}( \mathbf{h}_{k,1},  \ldots, \mathbf{h}_{k,N})$, $\mathbf{P}(k)=\mathrm{Diag}( \mathbf{p}_{k,1},  \ldots, \mathbf{p}_{k,N})$ and $\mathbf{Z}_{k}=\mathrm{Diag}((Z_{k,1}\hspace{-0.1cm}+\hspace{-0.1cm}J_{k,1}\hspace{-0.1cm}+\hspace{-0.1cm}N_{0}),\ldots,(Z_{k,N}\hspace{-0.1cm}+\hspace{-0.1cm}J_{k,N}\hspace{-0.1cm}+\hspace{-0.1cm}N_{0}) )$. Under the assumption of carrier non-cooperative processing, the mutual information at IU-$k$ is expressed as $I_{k}=\log|\mathbf{I}+\mathbf{Z}_{k}^{-1}\mathbf{H}^{H}_{k}\mathbf{P}(k)\mathbf{P}(k)^{H}\mathbf{H}_{k}|$ \cite{Palomar_2003, Cover_1991}. 

In order to obtain the optimal precoders that maximize the mutual information, we make use of the mutual information - Mean Square Error (MSE) relations. We note that in addition to the numerous works, such approach is taken for designing the optimal precoders for MC multi-antenna systems in \cite{Palomar_2003} and RSMA in MISO BC in \cite{Joudeh_2016}. We first obtain the optimal receive filter, $g_{k,n}$, that minimizes the Mean Square Error (MSE) \mbox{$\mathbb{E}\left\lbrace|\epsilon_{k,n}|^{2}\right\rbrace=\mathbb{E}\left\lbrace|g_{k,n}y_{k,n}-x_{k,n}|^{2} \right\rbrace$}, $\forall k \in \mathcal{K}$, $\forall n \in \mathcal{S}$. It is well known that the solution is given by a Minimum MSE (MMSE) filter 
\begin{align}
	g^{\mathrm{opt}}_{k,n}=\mathbf{p}^{H}_{k,n}\mathbf{h}_{k,n}\left( |\mathbf{h}_{k,n}^{H}\mathbf{p}_{k,n}|^{2}+Z_{k,n}+J_{k,n}+N_{0}\right) ^{-1}.
	\label{eqn:mmse}
\end{align}
The resulting MSE is written as
\begin{align}
	\epsilon^{\mathrm{opt}}_{k,n}\hspace{-0.1cm}=\hspace{-0.1cm}\left( |\mathbf{h}_{k,n}^{H}\mathbf{p}_{k,n}|^{2}\hspace{-0.1cm}+\hspace{-0.1cm}Z_{k,n}\hspace{-0.1cm}+\hspace{-0.1cm}J_{k,n}\hspace{-0.1cm}+\hspace{-0.1cm}N_{0}\right) ^{-1}\hspace{-0.1cm}(Z_{k,n}\hspace{-0.1cm}+\hspace{-0.1cm}J_{k,n}\hspace{-0.1cm}+\hspace{-0.1cm}N_{0}).
	\label{eqn:mse}
\end{align}
The mutual information-MSE relation is given by 
\begin{align}
	I_{k}=-\log|\mathbf{E}_{k}|,
	\label{eqn:mi}
\end{align}
where $\mathbf{E}_{k}=\mathrm{Diag}(\epsilon^{\mathrm{opt}}_{k,1}, \ldots, \epsilon^{\mathrm{opt}}_{k,N} )$ \cite{Palomar_2003}. The expression \eqref{eqn:mi} can be expanded as
\begin{align}
	\hspace{-0.1cm}I_{k}\hspace{-0.1cm}=\hspace{-0.1cm}-\log\left( \prod_{n=1}^{N}\epsilon^{\mathrm{opt}}_{k,n}\right)\hspace{-0.06cm}=\hspace{-0.06cm}-\sum_{n=1}^{N}\log(\epsilon^{\mathrm{opt}}_{k,n})\hspace{-0.06cm}=\hspace{-0.06cm}\sum_{n=1}^{N}I_{k,n}. \label{eqn:mutual_private}
\end{align}

In the context of conventional RSMA, \eqref{eqn:mutual_private} represents the mutual information for the private stream of SU-$k$. Similarly, the mutual information for the common stream at SU-$k$ is written as
\begin{align}
	I_{c,k}=-\sum_{n=1}^{N}\log(\epsilon^{\mathrm{opt}}_{c,k,n})=\sum_{n=1}^{N}I_{c,k,n}, \label{eqn:mutual_common}
\end{align}
where $\epsilon^{\mathrm{opt}}_{c,k,n}$ is obtained by replacing $\mathbf{p}_{k,n}$ and $Z_{k,n}$ in \eqref{eqn:mse} by $\mathbf{p}_{c,n}$ and $Z_{c,k,n}$, respectively. 
According to RSMA framework, the common stream should be decodable by all SUs in the system. Such requirement calls for a constraint on the mutual information of the common stream at subcarrier-$n$ as\footnote{Note that by carrier cooperative processing, the decodability can be guaranteed over $I_{c,k}$ instead of $I_{c,k,n}$ with proper coding methods \cite{Raleigh_1998}.}
\begin{align}
	I_{c,n}=\min_{k \in \mathcal{K}} I_{c,k,n}.
	\label{eqn:min_common}
\end{align}

In this work, we consider the ergodic mutual information $\mathbb{E}_{\mathbf{H}_{k}}\left\lbrace I_{c,k} \right\rbrace$ and $\mathbb{E}_{\mathbf{H}_{k}}\left\lbrace I_{k} \right\rbrace$ to investigate the average performance achieved under imperfect CSIT and over different CSIT realizations. Under carrier non-cooperative processing, we write $\mathbb{E}_{\mathbf{H}_{k}}\left\lbrace I_{c,k} \right\rbrace=\sum_{n=1}^{N}\mathbb{E}_{\mathbf{h}_{k,n}}\left\lbrace I_{c,k,n} \right\rbrace$ and $\mathbb{E}_{\mathbf{H}_{k}}\left\lbrace I_{k} \right\rbrace=\sum_{n=1}^{N}\mathbb{E}_{\mathbf{h}_{k,n}}\left\lbrace I_{k,n} \right\rbrace$.
Defining the average mutual information $\bar{I}_{c,k,n}= \mathbb{E}_{\mathbf{h}_{k,n}|\widehat{\mathbf{h}}_{k,n}}\left\lbrace I_{c,k,n}\right\rbrace$
and $\bar{I}_{k,n}= \mathbb{E}_{\mathbf{h}_{k,n}|\widehat{\mathbf{h}}_{k,n}}\left\lbrace I_{k,n}\right\rbrace$, it can be shown that $\mathbb{E}_{\mathbf{h}_{k,n}}\left\lbrace I_{c,k,n} \right\rbrace=\mathbb{E}_{\widehat{\mathbf{h}}_{k,n}}\left\lbrace \bar{I}_{c,k,n} \right\rbrace$ and $\mathbb{E}_{\mathbf{h}_{k,n}}\left\lbrace I_{k,n} \right\rbrace=\mathbb{E}_{\widehat{\mathbf{h}}_{k,n}}\left\lbrace \bar{I}_{k,n} \right\rbrace$ \cite{Joudeh_2016}. The constraint on the common stream in \eqref{eqn:min_common} is expressed in terms of average mutual information as $\bar{I}_{c,n}=\min_{k \in \mathcal{K}} \bar{I}_{c,k,n}$. We further define the rate portions $\bar{C}_{k,n}$, such that $\sum_{k \in \mathcal{K}}\bar{C}_{k,n}=\bar{I}_{c,n}$.

The average mutual information expressions will be used in the objective function of our problem formulation. Next, we determine the jamming and interference power constraints for the problem under statistical and imperfect CSIT, respectively.
\subsection{Jamming Power Constraint}
We aim to design a system that can focus a certain amount of power to pilot subcarriers of AUs. For this purpose, we consider the average power focused on subcarrier-$n$ of AU-$l$ \cite{Xing_2013}. Accordingly, we write
\begin{align}
	 &\hspace{-0.3cm}\bar{\Lambda}_{l,n}\triangleq\mathbb{E}\hspace{-0.05cm}\left\lbrace\hspace{-0.05cm}|\mathbf{g}_{l,n}^{H}\mathbf{p}_{c,n}|^{2}\hspace{-0.1cm}+\hspace{-0.1cm}\sum_{k \in \mathcal{K}}\hspace{-0.1cm}|\mathbf{g}_{l,n}^{H}\mathbf{p}_{k,n}|^{2}\hspace{-0.1cm}+\hspace{-0.1cm}\sum_{l^{\prime} \in \mathcal{L}}\hspace{-0.1cm}|\mathbf{g}_{l,n}^{H}\mathbf{f}_{l^{\prime},n}|^{2}\hspace{-0.05cm}\right\rbrace  \nonumber \\
	&\hspace{-0.3cm}=\mathbf{p}_{c,n}^{H}\mathbf{R}_{l,n}\mathbf{p}_{c,n}\hspace{-0.1cm}+\hspace{-0.1cm}\sum_{k \in \mathcal{K}}\mathbf{p}_{k,n}^{H}\mathbf{R}_{l,n}\mathbf{p}_{k,n}
	\hspace{-0.1cm}+\hspace{-0.1cm}\sum_{l^{\prime} \in \mathcal{L}}\mathbf{f}_{l^{\prime},n}^{H}\mathbf{R}_{l,n}\mathbf{f}_{l^{\prime},n}. 
	\label{eqn:average_power}
\end{align}
\subsection{Interference Power Constraint}
\label{sec:int_thr}
We consider the average total interference on all antennas of any given PU under imperfect CSIT. 
Accordingly, the conditional mean of the interference power on subcarrier-$n$ of PU-$m$ for a given $\wh{\mathbf{M}}_{m,n}$ is obtained as in \eqref{eqn:int_const} with
\begin{align}
 \mathbf{\Phi}_{m,n}=\hspace{-0.1cm}(1\minus\sigma_{pe}^{2}) \wh{\mathbf{M}}_{m,n}\wh{\mathbf{M}}_{m,n}^{H}\hspace{-0.1cm}+\hspace{-0.1cm}\sigma_{pe}^{2}N_{r,m}\mathbf{I}_{N_{t}}\hspace{-0.07cm}.
 \label{eqn:phi}
\end{align}

\begin{table*}
	\begin{align}
		\bar{\Psi}_{m,n}&\triangleq\mathbb{E}\left\lbrace||\mathbf{M}_{m,n}^{H}\mathbf{p}_{c,n}||^{2}+\sum_{k \in \mathcal{K}}||\mathbf{M}_{m,n}^{H}\mathbf{p}_{k,n}||^{2}+\sum_{l \in \mathcal{L}}||\mathbf{M}_{m,n}^{H}\mathbf{f}_{l,n}||^{2}\Bigg|\wh{\mathbf{M}}_{m,n}\right\rbrace \nonumber  \\
		&=(1\minus\sigma_{pe}^{2})\left[ \mathbf{p}_{c,n}^{H}\wh{\mathbf{M}}_{m,n}\wh{\mathbf{M}}_{m,n}^{H}\mathbf{p}_{c,n}\hspace{-0.12cm}+\hspace{-0.1cm}\sum_{k \in \mathcal{K}}\mathbf{p}_{k,n}^{H}\wh{\mathbf{M}}_{m,n}\wh{\mathbf{M}}_{m,n}^{H}\mathbf{p}_{k,n}+\sum_{l \in \mathcal{L}}\mathbf{f}_{l,n}^{H}\wh{\mathbf{M}}_{m,n}\wh{\mathbf{M}}_{m,n}^{H}\mathbf{f}_{l,n}\right]+\sigma_{pe}^{2}N_{r,m}\left[|\mathbf{p}_{c,n}|^{2}+\sum_{k \in \mathcal{K}} |\mathbf{p}_{k,n}|^{2}+\sum_{l \in \mathcal{L}}|\mathbf{f}_{l,n}|^{2} \right]   \nonumber\\
		&=\mathbf{p}_{c,n}^{H}\mathbf{\Phi}_{m,n}\mathbf{p}_{c,n}\hspace{-0.08cm}+\hspace{-0.1cm}\sum_{k \in \mathcal{K}}\hspace{-0.08cm}\mathbf{p}_{k,n}^{H}\mathbf{\Phi}_{m,n}\mathbf{p}_{k,n}\hspace{-0.08cm}+\hspace{-0.1cm}\sum_{l \in \mathcal{L}}\hspace{-0.08cm}\mathbf{f}_{l,n}^{H}\mathbf{\Phi}_{m,n}\mathbf{f}_{l,n}.
		\label{eqn:int_const}
	\end{align} 
	\vspace{-0.2cm}
	\hrule
	\vspace{-0.7cm}
\end{table*}
\subsection{Optimization Problem}
Using the average mutual information expressions for the objective function, and \eqref{eqn:average_power} and \eqref{eqn:int_const} as constraints, we formulate the optimization problem for RSMA as
\begin{subequations}
	\begin{alignat}{3}
		\hspace{-0.4cm}\max_{\mathbf{P},\mathbf{F},\bar{\mathbf{c}} }&     \ \ \sum_{k \in \mathcal{K}}\sum_{n \in \mathcal{S}}( \bar{C}_{k,n}+\bar{I}_{k,n})       \label{eqn:obj}   \\
		\text{s.t.}&  \ \  \sum_{k^{\prime} \in \mathcal{K}}\bar{C}_{k^{\prime},n} \leq \bar{I}_{c,k,n}, \quad \forall n \in \mathcal{S},  \forall k \in \mathcal{K} \label{eqn:common_rate_1} \\
		&\ \  \bar{\Lambda}_{l,n} \geq J^{thr}_{l,n},  \quad \forall n \in \mathcal{S}_{p,l}, \ \forall l \in \mathcal{L} \label{eqn:energy_1} \\
		&\ \ \bar{\Psi}_{m,n} \leq I^{thr}_{m,n},  \quad \forall n \in \mathcal{S}, \ \forall m \in \mathcal{M} \label{eqn:interference_1} \\
		& \ \  \sum_{n \in \mathcal{S}} \mathrm{tr}(\mathbf{P}_{n}\mathbf{P}_{n}^ {H}) + \mathrm{tr}(\mathbf{F}_{n}\mathbf{F}_{n}^ {H}) \leq \bar{P}_{t},  \label{eqn:total_power_1} \\
		& \ \  \frac{1}{N}\sum_{n \in \mathcal{S}}( \bar{C}_{k,n}+\bar{I}_{k,n}) \geq R_{th}, \quad \forall k \in \mathcal{K},   \label{eqn:QoS} 
	\end{alignat}
	\label{eqn:problem1}
\end{subequations}
\hspace{-0.15cm}where $\bar{\mathbf{c}}=[\bar{C}_{1,1}, \ldots, \bar{C}_{K,1}, \ldots, \bar{C}_{1,N}, \ldots, \bar{C}_{K,N}]$, $J^{thr}_{l,n}$ is the minimum threshold for the jamming energy focused on AU-$l$ at subcarrier-$n$, $I^{thr}_{m,n}$ is the maximum threshold for the interference at IU-$m$ at subcarrier-$n$, and $R_{th}$ is the minimum rate constraint to guarantee a minimum Quality-of-Service (QoS) for each user. 

\hspace{0.2cm}As mentioned in Section~\ref{sec:ergodic_mutual}, rate-MSE transformations as in \eqref{eqn:mutual_private} and \eqref{eqn:mutual_common} have been used in \cite{Joudeh_2016} without an MC waveform to transform the non-convex sum-rate maximization problem for RSMA into a convex one. Following the approach in \cite{Joudeh_2016}, we define the augmented weighted MSEs (WMSEs) as
\begin{align}
	\xi_{c,k,n}&=\omega_{c,k,n}\epsilon_{c,k,n}-\log_{2}(\omega_{c,k,n}), \nonumber  \\
	\xi_{k,n}&=\omega_{k,n}\epsilon_{k,n}-\log_{2}(\omega_{k,n}), 
	\label{eqn:wmse}
\end{align}
where $\omega_{c,k,n}$ and $\omega_{k,n}$ are the weights for the MSEs of the common and private streams at user-$k$ and subcarrier-$n$. 
It can be shown that the optimum weights are given by $\omega^{opt}_{c,k,n}=(\epsilon^{\mathrm{opt}}_{c,k,n})^{-1}$ and $\omega^{opt}_{k,n}=(\epsilon^{\mathrm{opt}}_{k,n})^{-1}$, for which the MSE-mutual information relations are obtained as
$\xi^{\mathrm{opt}}_{k,n}=1\minus I_{k,n}$ and $\xi^{\mathrm{opt}}_{c,k,n}=1\minus I_{c,k,n}$. Accordingly, the average augmented MSE-rate transformations are written as \cite{Joudeh_2016} 
\begin{align}
	\bar{\xi}^{\mathrm{opt}}_{c,k,n}=1\minus \bar{I}_{c,k,n}, \ \ \bar{\xi}^{\mathrm{opt}}_{k,n}=1\minus \bar{I}_{k,n}, \ \forall n \in \mathcal{S}.
	\label{eqn:augMSErate}
\end{align}
Using the average augmented MSEs in \eqref{eqn:augMSErate}, we transform the problem \eqref{eqn:problem1} as 
\begin{subequations}
	\begin{alignat}{3}
		\min_{\mathbf{P},\mathbf{F},\boldsymbol{\omega},\mathbf{g},\bar{\mathbf{x}}}&  \sum_{n \in \mathcal{S}}\sum_{k \in \mathcal{K}}(\bar{X}_{k,n}+\bar{\xi}_{k,n})      \label{eqn:obj_f}   \\
		\text{s.t.}&    \ \  1+\sum_{k^{\prime} \in \mathcal{K}}\bar{X}_{k^{\prime},n} \geq \bar{\xi}_{c,k,n}, \  \forall n \in \mathcal{S}, \forall k \in \mathcal{K},  \label{eqn:common_rate_1_f} \\
		& \ \  \frac{1}{N}\sum_{n \in \mathcal{S}}( \bar{X}_{k,n}+\bar{\xi}_{k,n}) \leq 1-R_{th}, \ \forall k \in \mathcal{K},   \label{eqn:QoS_2} \\
		&\ \eqref{eqn:energy_1}, \ \eqref{eqn:interference_1},\ \eqref{eqn:total_power_1},   
	\end{alignat}
	\label{eqn:problem2_final_nonconvex}
\end{subequations}
\hspace{-0.2cm}where 
\mbox{$\mathbf{x}=\left[\bar{X}_{1,1},\ldots,\bar{X}_{K,1}, \ldots, \bar{X}_{1,N},\ldots,\bar{X}_{K,N}\right]^{T} $}, $\bar{X}_{k,n}=-\bar{C}_{k,n}$, 
$\boldsymbol{\omega}_{n}=[\omega_{c,1,n}, \omega_{1,n} \ldots, \omega_{c,K,n}, \omega_{K,n}]^{T}$, $\boldsymbol{\omega}=[\boldsymbol{\omega}^{T}_{1}, \ldots, \boldsymbol{\omega}^{T}_{N}]^{T}$, $\mathbf{g}_{n}=[g_{c,1,n}, g_{1,n}, \ldots, g_{c,K,n}, g_{K,n}]^{T}$ and 
$\mathbf{g}=[\mathbf{g}^{T}_{1}, \ldots, \mathbf{g}^{T}_{N}]^{T}$. 
For the sake of brevity, we skip the detailed derivations to transform the stochastic problem (due to average augmented MSEs) into a deterministic form by the Sample Average Approximation (SAA) and refer the interested reader to \cite{Joudeh_2016}. 

The resulting problem formulation is non-convex with respect to $\left\lbrace \mathbf{P},\mathbf{F},\boldsymbol{\omega},\mathbf{g}\right\rbrace $ and also due to 
the constraint \eqref{eqn:energy_1}, in which the convex function $\bar{\Lambda}_{l,n}$ is constrained to a lower bound. In the next section, we propose an algorithm to solve the non-convex problem.

\section{Proposed Algorithm}
\label{sec:proposed}
In this section, we describe an AO-ADMM-based algorithm to solve the non-convex problem \eqref{eqn:problem2_final_nonconvex}.
ADMM is an algorithm that benefits from the decomposability of dual ascent method with the convergence properties of the method of multipliers \cite{boyd_2011}. We employ ADMM to split \eqref{eqn:problem2_final_nonconvex} into a minimization and a projection problem.
We define the vectors $\mathbf{v}=[\mathbf{v}_{1}^{T},\mathbf{v}_{2}^{T}\ldots, \mathbf{v}_{N}^{T}]^{T}$ and $\mathbf{v}_{n}=[X_{1,n}, \ldots, X_{K,n}, \mathrm{vec}(\mathbf{P}_{n})^{T}, \mathrm{vec}(\mathbf{F}_{n})^{T}]^{T}$. 
We reformulate the problem \eqref{eqn:problem2_final_nonconvex} according to the ADMM framework as 
\begin{subequations}
	\begin{alignat}{3}
		\min_{\mathbf{v},\mathbf{u}}&   \quad  f(\mathbf{v})+\Pi(\mathbf{u})      \label{eqn:obj_3}   \\
		\text{s.t.}&  \quad  \mathbf{v}-\mathbf{u}=\mathbf{0},
	\end{alignat}
	\label{eqn:problem_admm}
\end{subequations}
\hspace{-0.2cm}where 
$f(\mathbf{v})=\sum_{n \in \mathcal{S}}(\bar{\xi}_{c,n}+\sum_{k \in \mathcal{K}} \bar{\xi}_{k,n})$ and the vectors $\mathbf{u}_{n}$, $\forall n \in \mathcal{N}$, and $\mathbf{u}=[\mathbf{u}_{1}^{T},\mathbf{u}_{2}^{T}\ldots, \mathbf{u}_{N}^{T}]^{T}$ are introduced to split the problem according to the ADMM procedure. 
The function $\Pi(\mathbf{v})$ is the indicator function that performs projection on the domain $\mathcal{D}$ defined by the constraints \eqref{eqn:common_rate_1_f}, \eqref{eqn:QoS_2}, \eqref{eqn:energy_1},  \eqref{eqn:interference_1}, and \eqref{eqn:total_power_1}, {\sl i.e.}, $\Pi(\mathbf{u})=0$ if $\mathbf{u} \in \mathcal{D}$ and $\Pi(\mathbf{u})=\infty$ if $\mathbf{u} \notin \mathcal{D}$. 
We also define the real-valued vectors \mbox{$\mathbf{v}_{r,n}=[\mathfrak{R}(\mathbf{v}_{n}^{T}), \mathfrak{I}(\mathbf{v}_{n}^{T})]^{T}$}, \mbox{$\mathbf{u}_{r,n}=[\mathfrak{R}(\mathbf{u}_{n}^{T}), \mathfrak{I}(\mathbf{u}_{n}^{T})]^{T}$}, \mbox{$\mathbf{v}_{r}=[\mathbf{v}_{r,1}^{T},\mathbf{v}_{r,2}^{T}\ldots, \mathbf{v}_{r,N}^{T}]^{T}$} and \mbox{$\mathbf{u}_{r}=[\mathbf{u}_{r,1}^{T},\mathbf{u}_{r,2}^{T}\ldots, \mathbf{u}_{r,N}^{T}]^{T}$} 
in accordance with the ADMM framework, where the functions $\mathfrak{R}(.)$ and $\mathfrak{I}(.)$ return the real and imaginary parts of their inputs, respectively. 
 \begin{figure}[t!]
 	\removelatexerror
 	\begin{algorithm}[H]
 		\caption{AO-ADMM-Based Algorithm}
 		\label{alg:admm}
 		$t \gets 0$, $i \gets 0$, $\mathbf{v}_{r}^{0}$, $\mathbf{u}_{r}^{0}$,  $\mathbf{w}_{r}^{0}$, $\hat{\mathbf{u}}_{r}\gets\mathbf{u}_{r}^{0}$ \\
 		\While{$|\mathrm{SR}^{i}-\mathrm{SR}^{i-1}|> \epsilon_{r}$}{
 			$\boldsymbol{\omega}^{i} \gets$ updateWeights($\hat{\mathbf{u}}_{r}$) \\
 			$\boldsymbol{g}^{i} \gets$ updateFilters($\hat{\mathbf{u}}_{r}$) \\
 			\While{$\sum_{n\in\mathcal{S}}\hspace{-0.08cm}||\mathbf{r}_{r,n}^{t}||\hspace{-0.08cm}>\hspace{-0.08cm}\epsilon_{a}$ $\mathrm{\mathbf{or}}$ $\sum_{n\in\mathcal{S}}\hspace{-0.08cm}||\mathbf{q}_{r,n}^{t}||\hspace{-0.08cm}>\hspace{-0.08cm}\epsilon_{a}$}{
 				$\mathbf{v}_{r}^{t+1} \gets \argmin_{\mathbf{v}_{r}}(f(\mathbf{v}_{r},\boldsymbol{\omega}^{i},\mathbf{g}^{i})+\frac{\zeta}{2}\sum_{n\in\mathcal{S}}||\mathbf{v}_{r,n}-\mathbf{u}^{t}_{r,n}+\mathbf{w}^{t}_{r,n}||^{2} )$ via interior-point methods\\
 				$\mathbf{u}^{t+1}_{r} \gets \argmin_{\mathbf{u}_{r}}(\Pi(\mathbf{u}_{r}) +\frac{\zeta}{2}\sum_{n\in\mathcal{S}}||\mathbf{v}^{t+1}_{r,n}-\mathbf{u}_{r,n}+\mathbf{w}^{t}_{r,n}||^{2} )$ via SDR method\\
 				$\mathbf{w}^{t+1}_{r,n}=\mathbf{w}^{t}_{r,n}+\mathbf{v}^{t+1}_{r,n}-\mathbf{u}^{t+1}_{r,n}, \quad \forall n \in \mathcal{N}$\\
 				$\mathbf{r}^{t+1}_{r,n} \gets  \mathbf{v}^{t+1}_{r,n}-\mathbf{u}^{t+1}_{r,n}, \quad \forall n \in \mathcal{N}$\\
 				$\mathbf{q}^{t+1}_{r,n} \gets \mathbf{u}^{t+1}_{r,n}-\mathbf{u}^{t}_{r,n}, \quad \forall n \in \mathcal{N}$\\
 				$t \gets t + 1$\\
 			}
 			$\hat{\mathbf{u}}_{r} \gets \mathbf{u}_{r}^{t}$, \ $\mathbf{u}_{r}^{0} \gets \mathbf{u}_{r}^{t}$  \\
 			 $\mathrm{SR}^{i+1} \gets$ updateSR($\hat{\mathbf{u}}_{r}$) \\
 			$t \gets 0$, \ $i \gets i + 1$\\
 		}
 		\Return $\hat{\mathbf{u}}_{r}$
 	\end{algorithm}
 	\vspace{-0.5cm}
 \end{figure}

The augmented Lagrangian function for the optimization problem \eqref{eqn:problem_admm} is written as 
\begin{align}
	\mathcal{L}_{\zeta}&(\mathbf{v}_{r},\mathbf{u}_{r},\mathbf{d}_{r})= f(\mathbf{v_{r}})+\Pi(\mathbf{u}_{r}) \nonumber \\
	&+\sum_{n \in \mathcal{S}} \mathbf{d}_{r,n}^{T}(\mathbf{v}_{r,n}-\mathbf{u}_{r,n})+\sum_{n \in \mathcal{S}} (\zeta/2)||\mathbf{v}_{r,n}-\mathbf{u}_{r,n}||^{2}_{2}, \nonumber
\end{align}
\noindent where $ \zeta > 0$ is called the penalty parameter, $\mathbf{d}_{r}=[\mathbf{d}_{r,1}^{T},\mathbf{d}_{r,2}^{T}\ldots, \mathbf{d}_{r,N}^{T}]^{T}$ and $\mathbf{d}_{r,n}=[\mathfrak{R}(\mathbf{d}_{n}^{T}), \mathfrak{I}(\mathbf{d}_{n}^{T})]^{T}$ and $\mathbf{d}_{n} \in \mathbb{C}^{(K+N_{t}(K+L+1))}$, $\forall n \in \mathcal{S}$ being the dual variables. 
The updates of iterative ADMM procedure can be written in the scaled form as
\begin{align}
		\hspace{-0.3cm}&\mathbf{v}^{t+1}_{r}\hspace{-0.1cm}=\hspace{-0.1cm}\argmin_{\mathbf{v}_{r}}(f(\mathbf{v}_{r}))\hspace{-0.1cm}+\hspace{-0.1cm}\frac{\zeta}{2}\hspace{-0.1cm}\sum_{n\in\mathcal{S}}||\mathbf{v}_{r,n}-\mathbf{u}^{t}_{r,n}+\mathbf{w}^{t}_{r,n}||^{2} )  \label{eqn:v_update} \\
		\hspace{-0.3cm}&\mathbf{u}^{t+1}_{r}\hspace{-0.1cm}=\hspace{-0.1cm}\argmin_{\mathbf{u}}(\Pi(\mathbf{u}_{r})\hspace{-0.1cm}+\hspace{-0.1cm}\frac{\zeta}{2}\hspace{-0.1cm}\sum_{n\in\mathcal{S}}||\mathbf{v}^{t+1}_{r,n}-\mathbf{u}_{r,n}+\mathbf{w}^{t}_{r,n}||^{2} )  \label{eqn:u_update}  \\
		\hspace{-0.3cm}&\mathbf{w}^{t+1}_{r,n}\hspace{-0.1cm}=\hspace{-0.1cm}\mathbf{w}^{t}_{r,n}+\mathbf{v}^{t+1}_{r,n}-\mathbf{u}^{t+1}_{r,n}, \quad n \in \mathcal{N}, \label{eqn:w_update}
\end{align}
where $\mathbf{w}_{r}=\mathbf{d}_{r}/\zeta$. The update step for $\mathbf{v}_{r}$ in \eqref{eqn:v_update} deals with the WMSE minimization problem. The update step for $\mathbf{u}_{r}$ in \eqref{eqn:u_update} involves projection onto the domain $\mathcal{D}$.

The AO-ADMM-based algorithm to solve the updates in \eqref{eqn:v_update}-\eqref{eqn:w_update} is given in Alg.~\ref{alg:admm}. 
The outer iterations of the AO algorithm serve to update the MSE weights $\boldsymbol{\omega}^{i}$ and equalizers $\mathbf{g}^{i}$ based on the precoders calculated by ADMM at iteration-$i$. By the calculated $\boldsymbol{\omega}^{i}$ and $\mathbf{g}^{i}$, the ADMM algorithm performs the update steps in \eqref{eqn:v_update} and \eqref{eqn:u_update}. Note that we include the terms $\boldsymbol{\omega}^{i}$ and $\mathbf{g}^{i}$ in the function representations $f(.)$ and $\Pi(.)$ to highlight the dependencies on the corresponding parameters.

The update step for $\mathbf{v}_{r}$ can be solved using interior-point methods.  We propose the Semi-Definite Relaxation (SDR) method for the update step for $\mathbf{u}_{r}$ to deal with the non-convex constraint \eqref{eqn:energy_1} \cite{ma_2010}. 
Consider the following formulation for the minimization step for $\mathbf{u}$ expressed in terms of the complex-valued parameters as
\begin{subequations}
	\begin{alignat}{3}
		\min_{\mathbf{u}}&     \quad  \sum_{n\in\mathcal{S}}||\mathbf{v}^{t+1}_{n}-\mathbf{u}_{n}+\mathbf{w}^{t}_{n}||^{2}         \\
		\text{s.t.}& \quad 1+\sum_{k^{\prime} \in \mathcal{K}}\mathbf{e}^{T}_{k^{\prime}}\mathbf{u}_{n}\geq  \bar{\xi}_{c,k,n},  \forall n \in \mathcal{S},  \forall k \in \mathcal{K} \label{eqn:u_upd_1}\\
		&\quad  \bar{\Lambda}_{l,n} \geq J^{thr}_{l,n},  \quad \forall n \in \mathcal{S}_{p,l}, \ \forall l \in \mathcal{L} \label{eqn:u_upd_2}\\
		&\quad   \bar{\Psi}_{m,n} \leq I^{thr}_{m,n},  \ \forall n \in \mathcal{S}, \ \forall m \in \mathcal{M} \label{eqn:u_upd_3}\\
		&\quad \sum_{n \in \mathcal{S}}\mathrm{tr}\left( \mathbf{D}_{s}\mathbf{u}_{n}(\mathbf{D}_{s}\mathbf{u}_{n})^{H}\right) \leq \bar{P}_{t},  \label{eqn:u_upd_4} \\
		& \ \  \frac{1}{N}\sum_{n \in \mathcal{S}}( \mathbf{e}^{T}_{k}\mathbf{u}_{n}+\bar{\xi}_{k,n}) \leq 1-R_{th}, \quad \forall k \in \mathcal{K},   \label{eqn:QoS_3} 
	\end{alignat}
	\label{eqn:problem_umin}
\end{subequations}
\hspace{-0.1cm}where $\mathbf{D}_{s}=[\mathbf{0}^{(K+L+1)N_{t} \times (1)}, \ \mathbf{I}_{(K+L+1)N_{t}}]$ and $\mathbf{e}_{k}$ is the $k$-th standard basis vector of length $N_{t}(K+L+1)+1$. The inhomogeneous Quadratically Constrained Quadratic
Program (QCQP) formulation \eqref{eqn:problem_umin} can be transformed into an equivalent homogeneous QCQP as demonstrated in \cite{ma_2010}, and the resulting problem can be solved by applying the SDR procedure.  
We omit the rest of the details on the application of the SDR procedure due to lack of space and refer the interested reader to \cite{ma_2010}.

\textit{Convergence:}
We discuss the convergence over the real-valued equivalent definitions of the functions. First, consider the inner problem in Alg.~\eqref{alg:admm}, 
\begin{subequations}
	\begin{alignat}{3}
		\min_{\mathbf{v}_{r}}&  \quad  f(\mathbf{v_{r}},\boldsymbol{\omega}^{i},\mathbf{g}^{i})  \nonumber \\
		\text{s.t.}& \quad  \mathbf{v}_{r} \in \mathcal{D}^{i}, \nonumber
	\end{alignat}
	\label{eqn:problem_admm_conv}
\end{subequations}
\hspace{-0.2cm}where $\mathcal{D}^{i}$ represents the problem domain in the $i$-th iteration of the AO algorithm. It can be shown that the function $f(\mathbf{v_{r}},\boldsymbol{\omega}^{i},\mathbf{g}^{i})$ is Lipschitz differentiable and $\mathcal{D}^{i}$ is a compact set. Then, the sequence
($\mathbf{v}^{t}_{r}$ , $\mathbf{u}^{t}_{r}$, $\mathbf{w}^{t}_{r}$) has at least one limit point, and each limit point is a stationary point of $\mathcal{L}_{\zeta}(\mathbf{v}_{r},\mathbf{u}_{r},\mathbf{w}_{r})$ for any sufficiently large $\zeta$ \cite[Cor. 2]{wang_2019}. This is also valid when the subproblems are solved inexactly with summable errors, which is a condition satisfied by the SDR method \cite{zuo_2010}.
Given the convergence of the ADMM part for the inner iterations, the convergence of the AO algorithm for the outer iterations follow from \cite[Prop. 1]{Joudeh_2016}.

\textit{Complexity:}
We calculate the complexity of the proposed algorithm at each iteration by calculating the complexities of the update steps for $\mathbf{v}_{r}$ and $\mathbf{u}_{r}$ .
\begin{enumerate}
	\item Update step for $\mathbf{v}_{r}$: The complexity of this step is equal to the complexity of interior-point methods for solving the problem \eqref{eqn:v_update}, which can be written in the equivalent QCQP form 
	\begin{subequations}
		\begin{alignat}{3}
			\min_{\mathbf{v}_{r}, t}&  \quad  f(\mathbf{v_{r}})+t  \nonumber \\
			\text{s.t.}& \quad  \frac{\zeta}{2}\hspace{-0.1cm}\sum_{n\in\mathcal{S}}||\mathbf{v}_{r,n}-\mathbf{u}^{t}_{r,n}+\mathbf{w}^{t}_{r,n}||^{2} \leq t, \nonumber
		\end{alignat}
	\end{subequations}
where $t$ is a slack variable. The complexity of solving such a problem by interior point methods is $\mathcal{O}((N(K+L+2)N_{t}+1)^{3})=\mathcal{O}((N(K+L+2)N_{t})^{3})$, with $\mathcal{O}(.)$ denoting the big O function \cite{nesterov_1994}.
	\item Update step for $\mathbf{u}_{r}$: The complexity of this step is equal to the complexity of SDR method, which is given by $\mathcal{O}((N(K+L+2)N_{t})^{4.5}\log(1/\epsilon))$ for given solution accuracy given a solution accuracy $\epsilon > 0$ \cite{ma_2010}.
\end{enumerate}
Consequently, the complexity of a single iteration of the proposed algorithm is dominated by that of SDR method, which is $\mathcal{O}((N(K+L+2)N_{t})^{4.5}\log(1/\epsilon))$.

\section{Feasibility Analysis and Threshold Selection}
\label{sec:threshold}
In this section, we discuss on how to set $J^{thr}_{l,n}$ and $I^{thr}_{m,n}$.
The performance metric for jamming in our system model is the focused energy on the pilot subcarriers of AUs. Our aim is to focus as much jamming power as possible on each pilot subcarrier of each AU under the total transmit power constraint. The minimum focused jamming power is determined by the threshold $J^{thr}_{l,n}$ in constraint \eqref{eqn:energy_1} of problem \eqref{eqn:problem1}. Therefore, the threshold $J^{thr}_{l,n}$ should be chosen carefully to focus the desired amount of jamming power while guaranteeing a feasible problem. 

Additionally, one can observe the constraints \eqref{eqn:energy_1} and \eqref{eqn:interference_1} in \eqref{eqn:problem1} are conflicting in the sense that the former imposes a lower bound on the focused power on an AU, while the latter imposes an upper bound on the interference power at a PU at the same pilot subcarrier. Depending on the chosen thresholds $J^{thr}_{l,n}$ and $I^{thr}_{m,n}$, there might occur cases where satisfying both constraints is impossible, leading to an infeasible problem. A simple but relevant example is when AU-$l$ and PU-$m$ are co-located, so that their channels are almost identical. In this case, setting $I^{thr}_{m,n}<J^{thr}_{l,n}$ returns an empty domain for the problem \eqref{eqn:problem2_final_nonconvex}, since the constraints \eqref{eqn:energy_1} and \eqref{eqn:interference_1} cannot be satisfied at the same time.

In this section, we first present a rule to set the jamming power threshold $J^{thr}_{l,n}$. Then, we perform a feasibility analysis over the domain of the problem \eqref{eqn:problem2_final_nonconvex} to determine the conditions and optimal values for $I^{thr}_{m,n}$ for given $J^{thr}_{l,n}$ and system settings. Finally, we propose a low-complexity algorithm to set $I^{thr}_{m,n}$ for given $J^{thr}_{l,n}$ and system settings, which guarantees a non-empty domain for the problem and achieves the optimal interference power threshold values under specific conditions found by the feasibility analysis. The feasibility analysis and the proposed threshold selection methods explained in this section guarantee that the resulting problem is feasible and robust against variations in the thresholds, in case the thresholds are varied within the bounds demonstrated in this section.
\subsection{Setting the Jamming Power Threshold}
In this section, we discuss how to set the jamming power threshold in the constraint \eqref{eqn:energy_1} for pilot subcarriers. It has been shown in \cite{Ohno_2002, negi_1998, Adireddy_2002} that equally spaced pilot insertion and uniform power allocation over all pilots is optimal for estimation and equalization of the channel frequency response at the receiver. 
Assuming the transmissions to AUs from their corresponding friendly units follow such optimal allocation, we set the power threshold by giving equal weights to all pilot subcarriers of the AUs. 
Considering the discussions above, we present a method to set $J^{thr}_{l,n}$ in Proposition 1 that can achieve the maximum focused jamming power while guaranteeing a feasible solution for problem \eqref{eqn:problem1} (assuming $I^{thr}_{m,n}$ is large enough).

\textit{Proposition 1:}  Consider the optimization problem \eqref{eqn:problem2_final_nonconvex} with the constraint \eqref{eqn:interference_1} excluded. The resulting problem is guaranteed to be feasible if
\begin{align}
	J^{thr}_{l,n}=\rho \frac{\bar{P}_{t}}{N_{p,l}L}\sigma^{max}_{l,n},  \quad \forall n \in \mathcal{S}_{p,l}, \ \forall l \in \mathcal{L},
	\label{eqn:jamm_thr}
\end{align} 
for any $0 \leq \rho \leq 1 $, where $\sigma^{max}_{l,n}$ is the largest eigenvalue of the matrix $\mathbf{R}_{l,n}$ and \mbox{$N_{p,l}=|\mathcal{S}_{p,l}|$} denotes the number of pilot subcarriers of AU-$l$. 

\textit{Proof:} Please see Appendix~\ref{sec:append_prop1}.  \hspace{2.33cm}$\blacksquare$

Proposition 1 states that in order to guarantee a feasible solution of the problem for a generic scenario (without any consideration of the overlapping in eigenvector directions of AU channels), the maximum value one can set as the jamming power threshold is $\frac{\bar{P}_{t}}{N_{p,l}L}\sigma^{max}_{l,n}$, $\forall n \in \mathcal{S}_{p,l}$, $\forall l \in \mathcal{L}$, under the assumption that all AUs are jammed with an equal share of transmit power and the jamming power allocated to an AU is distributed equally among its pilot subcarriers. 
The parameter $\rho$ in \eqref{eqn:jamm_thr} is used to adjust the strictness of the jamming power constraints. 
Note that the maximum threshold value stated above is attained by \eqref{eqn:jamm_thr} when $\rho=1$. If $\rho<1$, the jamming power constraints are relaxed, leading to a larger domain for the considered problem. 

\subsection{Feasibility Analysis and Setting the Interference Power Threshold}
\label{sec:feasibility_analysis}
In this section, we investigate the problem of determining $I^{thr}_{m,n}$, so that, the problem \eqref{eqn:problem1} is feasible for a given $J^{thr}_{l,n}$ determined according to Proposition 1. Our aim is to determine the optimum values for $I^{thr}_{m,n}$ and the conditions these values are attained under. In other words, we search for a $I^{thr}_{m,n}$ which is large enough to define a non-empty domain for given $J^{thr}_{l,n}$ but small enough to limit the interference to PUs as much as possible. Then, we propose a low-complexity algorithm that is used to determine the interference power thresholds $I^{thr}_{m,n}$ for given $J^{thr}_{l,n}$ and system parameters. We show that the proposed algorithm returns threshold values which guarantee a non-empty domain for problem \eqref{eqn:problem1} under any given scenario, and attain the optimal values found by the feasibility analysis when the necessary conditions are satisfied.

First, we introduce some useful definitions and state some properties of the involved parameters. 
The eigen decompositions of the positive semi-definite matrices \mbox{$\mathbf{R}$} and \mbox{$\mathbf{\Phi}$} are given as
\mbox{$	\mathbf{R}=\mathbf{U}\Sigma \mathbf{U}^{H}$} and  \mbox{$\mathbf{\Phi}=\mathbf{V}\Lambda \mathbf{V}^{H}$},
where $\mathbf{U} \in \mathbb{C}^{N_{t}\times N_{t}}$ and $\mathbf{V} \in \mathbb{C}^{N_{t}\times N_{t}}$ are unitary matrices, {\sl i.e.,}  \mbox{$\mathbf{U}^{H}\mathbf{U}=\mathbf{U}\mathbf{U}^{H}=\mathbf{I}_{N_{t}}$} and \mbox{$\mathbf{V}^{H}\mathbf{V}=\mathbf{V}\mathbf{V}^{H}=\mathbf{I}_{N_{t}}$}, and $\Sigma$ and $\Lambda$ are diagonal matrices containing the non-negative eigenvalues of the corresponding matrices in their diagonal elements. 
An alternative representation for the eigen decomposition is written as \mbox{$\mathbf{R}=\sum_{i=1}^{n_{g}}\sigma^{i} \mathbf{u}^{i}\left(\mathbf{u}^{i}\right)^{H} $} 
\mbox{$\mathbf{\Phi}=\sum_{i=1}^{n_{m}}\lambda^{i} \mathbf{v}^{i}\left(\mathbf{v}^{i}\right) ^{H}$}, where $\sigma^{i}$ and $\lambda^{i}$ are the non-zero diagonal elements of $\Sigma$ and $\Lambda$, and \mbox{$\mathbf{u}^{i} \in \mathbb{C}^{N_{t}}$} and $\mathbf{v}^{i}$ are the $i$-th columns of $\mathbf{U}$ and $\mathbf{V}$, respectively. The terms $n_{g}$ and $n_{m}$ denote the number of non-zero eigenvalues of the matrices $\mathbf{R}$ and $\mathbf{\Phi}$, respectively.
Note the useful property that is, the eigen decompositions and Singular Value Decompositions (SVDs) are equivalent for positive semi-definite matrices.
In the analysis, we assume that the eigenvalue representations satisfy \mbox{$\sigma^{1}\geq \sigma^{2}\geq \ldots \geq \sigma^{n_{g}}$} and \mbox{$\lambda^{1}\geq \lambda^{2}\geq \ldots \geq \lambda^{n_{m}}$}. 

Proposition 1 states that the jamming power threshold in \eqref{eqn:jamm_thr} can be attained assuming a portion $\rho$ of the total transmit power is allocated equally among $\mathbf{f}_{l,n}$, $\forall n \in N_{p,l}$ and $\forall l \in \mathcal{L}$. In the following analysis, we consider the relation between focused powers on a given PU and a given AU with the abovementioned assumption on a given pilot subcarrier, and drop the user and subcarrier indexes for simplicity.  
We consider the following problem formulation
\begin{subequations}
	\begin{alignat}{3}
		\min_{\mathbf{p}}&     \ 0   \\ 
		\text{s.t.}&\quad  \mathbf{f}\mathbf{\Phi}\mathbf{f}\leq I,    \\
		&\quad  \mathbf{f}\mathbf{R}\mathbf{f} \geq \rho \frac{\bar{P}_{t}\sigma^{max}}{N_{p}L},   \\
		&\quad ||\mathbf{f}||^{2} \leq  \frac{\bar{P}_{t}}{N_{p}L}.   
	\end{alignat}
	\label{eqn:problem_analysis2}
\end{subequations}
In order to obtain a tractable problem, we transform \eqref{eqn:problem_analysis2} into the equivalent form 
\begin{subequations}
	\begin{alignat}{3}
		\min_{\mathbf{S}}&   \quad 0  \label{eqn:obj_sdp} \\
		\text{s.t.}&\quad  \mathrm{tr}\left(\mathbf{S} \mathbf{\Phi}   \right)   \leq  I, \label{eqn:int_sdp}   \\
		&\quad   \mathrm{tr}\left(\mathbf{S} \mathbf{R}   \right) \geq \rho \frac{\bar{P}_{t}}{N_{p}L}\sigma^{max},  \label{eqn:jamm_sdp} \\
		&\quad \mathrm{tr}\left(\mathbf{S}\right)  \leq  \frac{\bar{P}_{t}}{N_{p}L}  \label{eqn:power_sdp} , \\
		&\quad  \mathbf{S} \succcurlyeq 0 \label{eqn:s_sdp}, \\
		&\quad \mathrm{rank}\left(\mathbf{S}\right)=1, 
	\end{alignat}
	\label{eqn:problem_analysis_sdp}
\end{subequations}
\hspace{-0.15cm}where $\mathbf{S}=\mathbf{f}\mathbf{f}^{H}$. In the rest of the analysis, we do not consider the non-convex rank constraint in accordance with the SDR approach \cite{zuo_2010}.
The Lagrangian for the resulting convex problem is written as
\begin{align}
	&\mathcal{L}\left(\mathbf{S}, \alpha, \beta, \gamma, \mathbf{\Upsilon}\right)\hspace{-0.08cm}=\hspace{-0.08cm}\alpha\left(\hspace{-0.05cm}\mathrm{tr}\hspace{-0.08cm}\left(\mathbf{S} \mathbf{\Phi} \hspace{-0.05cm}\right)\hspace{-0.08cm}-\hspace{-0.08cm}I\right)\hspace{-0.08cm}-\hspace{-0.08cm} \mathrm{tr}\hspace{-0.08cm}\left(\mathbf{\Upsilon}  \mathbf{S}\right) \nonumber \\
	&\hspace{0.8cm}-\hspace{-0.08cm}\beta\hspace{-0.08cm}\left(\hspace{-0.08cm}\mathrm{tr}\left(\mathbf{S} \mathbf{R}\right)\hspace{-0.08cm}-\hspace{-0.08cm}\rho\frac{\bar{P}_{t}}{N_{p}L}\sigma^{max}\hspace{-0.05cm}\right)  
	\hspace{-0.08cm}+\hspace{-0.08cm}\gamma\hspace{-0.08cm}\left(\hspace{-0.08cm} \mathrm{tr}\left(\mathbf{S} \right)\hspace{-0.08cm}-\hspace{-0.08cm}\frac{\bar{P}_{t}}{N_{p}L}\hspace{-0.05cm}\right), \nonumber   
\end{align} 
where $\alpha$, $\beta$, $\gamma$ and $\mathbf{\Upsilon}$ are the Lagrangian multipliers. The Karush-Kuhn-Tucker (KKT) conditions are written as
\begin{subequations}
	\begin{alignat}{3}
	&\alpha^{*}\mathbf{\Phi}-\beta^{*}\mathbf{R}_{g}+\gamma^{*}\mathbf{I}_{N_{t}}=\mathbf{\Upsilon}^{*}, \label{eqn:kkt_0}\\ 
	&\alpha^{*}\left( \mathrm{tr}\left(\mathbf{S}^{*} \mathbf{\Phi}   \right)-I^{*}\right)=0, \label{eqn:kkt_1_0}\\
	&\beta^{*}\left( \mathrm{tr}\left(\mathbf{S}^{*} \mathbf{R}\right)-\rho\frac{\bar{P}_{t}}{N_{p}L}\sigma^{max} \right)=0, \label{eqn:kkt_1}\\ 
	&\gamma^{*}\left( \mathrm{tr}\left(\mathbf{S}^{*} \right)-\frac{\bar{P}_{t}}{N_{p}L}\right)=0, \label{eqn:kkt_2} \\
	&\mathrm{tr}\left(\mathbf{\Upsilon}^{*}\mathbf{S}^{*} \right) = 0 , \label{eqn:kkt_3} \\
	&\alpha^{*}, \beta^{*}, \gamma^{*}\geq 0, \mathbf{\Upsilon}^{*} \succcurlyeq 0, \label{eqn:kkt_4}\\
	& \eqref{eqn:int_sdp}, \eqref{eqn:jamm_sdp}, \eqref{eqn:power_sdp}, \eqref{eqn:s_sdp} \label{eqn:kkt_5}.
	\end{alignat}
\end{subequations}

The expression \eqref{eqn:kkt_0} is the stationarity constraints, the expressions \eqref{eqn:kkt_1}-\eqref{eqn:kkt_3} are the complementary slackness constraints, \eqref{eqn:kkt_4} are the dual feasibility and \eqref{eqn:kkt_5} are the primal feasibility constraints. 
We aim to consider the scenarios in which the interference constraint is active. In other words, we desire the interference to be limited by the interference power threshold rather than the precoder power constraint. Such an assumption also leads to distribution of more power to the data subcarriers, potentially leading to an increased sum-rate.
Therefore, we perform the analysis for the cases with  $\alpha^{*}>0$, $\gamma^{*}=0$, and we set \mbox{$\mathrm{tr}\left(\mathbf{S}^{*} \mathbf{\Phi}\right)=I^{*}$} and \mbox{$ \mathrm{tr}\left(\mathbf{S}^{*} \right)<\frac{\bar{P}_{t}}{N_{p}L}$} in the rest of the analysis.

First, we consider the case where constraint \eqref{eqn:kkt_1} is active and  $\beta^{*}>0$. 
The stationarity constraint \eqref{eqn:kkt_0} gives \mbox{$\mathbf{\Phi}-\beta^{*}\mathbf{R}=\mathbf{\Upsilon}^{*}$}, for some \mbox{$\mathbf{\Upsilon}^{*} \succcurlyeq 0$} by the dual feasibility constraint \eqref{eqn:kkt_4}. 
Then, the existence of $\beta^{*}$ depends on the characteristics of $\mathbf{\Phi}$ and $\mathbf{R}$ and its calculation is not straightforward in general. 
In case such a $\beta^{*}$ value can be obtained, we get \mbox{$\mathrm{tr}\left(\mathbf{S}^{*}\left(\mathbf{\Phi}-\beta^{*}\mathbf{R}\right)\right) \geq 0$}, resulting in
\begin{align}
	I^{*}\hspace{-0.1cm}=\hspace{-0.1cm}\mathrm{tr}\left(\mathbf{S}^{*}\mathbf{\Phi}\right)\hspace{-0.1cm}\geq\hspace{-0.1cm} \beta^{*}\mathrm{tr}\left(\mathbf{S}^{*}\mathbf{R}\right)=\beta^{*}\rho\frac{\bar{P}_{t}\sigma^{max}}{N_{p}L}, 
	\label{eqn:thr1}
\end{align}
where the second equality comes from the active constraint \eqref{eqn:kkt_1}. A special case for which $I^{*}$ can be calculated is given in Proposition 2. 

\textit{Proposition 2:} If $\mathbf{R}=\mathbf{I}_{N_{t}}$, $N_{t}>N_{r}$ and $0 \leq\rho<1$, a solution $\mathbf{S}^{*}=\mathbf{f}^{*}(\mathbf{f}^{*})^{H}$ exists for some $\mathbf{f}^{*} \in \mathbb{C}^N_{t}$ satisfying \mbox{$\mathrm{tr}\left(\mathbf{S}^{*} \right)<\frac{\bar{P}_{t}}{N_{p}L}$}, which gives
\begin{align}
	I^{*}\hspace{-0.1cm}=\hspace{-0.1cm}\mathrm{tr}\left(\mathbf{S}^{*}\mathbf{\Phi}\right)=\rho\frac{\sigma_{pe}^{2}N_{r,m}\bar{P}_{t}}{N_{p}L}.  
	\label{eqn:thr_special}
\end{align}

\textit{Proof:}  Please see Appendix~\ref{sec:append_prop2}.  \hspace{2.93cm}$\blacksquare$

We can conclude that the interference power threshold for the considered case is found for specific AU and PU channels and difficult to calculate for generic cases. 

Next, we consider the case where the constraint \eqref{eqn:kkt_1} is inactive, so that $\beta^{*}=0$. For the case \mbox{$\mathrm{tr}\left(\mathbf{S}^{*} \right)<\frac{\bar{P}_{t}}{N_{p}L}$}, Proposition 1 states that this particular case can occur only for $\rho < 1$. The stationarity constraint \eqref{eqn:kkt_0} is satisfied since \mbox{$\mathbf{\Phi}=\mathbf{\Upsilon}^{*}$}. Using the complementary slackness constraint \eqref{eqn:kkt_3}, we obtain 
\begin{align}
	\mathrm{tr}\hspace{-0.08cm}\left(\mathbf{\Upsilon}^{*}\mathbf{S}^{*}\hspace{-0.03cm} \right)\hspace{-0.08cm}=\hspace{-0.08cm}\mathrm{tr}\hspace{-0.08cm}\left( \mathbf{\Phi}\mathbf{S}^{*}\hspace{-0.03cm}\right)\hspace{-0.08cm}=\hspace{-0.08cm}\mathrm{tr}\hspace{-0.08cm}\left(\mathbf{S}^{*} \mathbf{\Phi}\right)\hspace{-0.08cm}=\hspace{-0.08cm}I^{*}\hspace{-0.13cm}=\hspace{-0.08cm}0.
	\label{eqn:thr2}
\end{align}
However, due to the imperfect CSIT assumption and $\mathrm{tr}\hspace{-0.08cm}\left(\mathbf{S}^{*} \mathbf{\Phi}\right)=(1\minus\sigma_{pe}^{2})\mathrm{tr}\hspace{-0.08cm}\left(\mathbf{S}^{*}  \wh{\mathbf{M}}\wh{\mathbf{M}}^{H}\right)+\sigma_{pe}^{2}N_{r,m}\mathrm{tr}\hspace{-0.08cm}\left(\mathbf{S}^{*} \right)$, the interference threshold cannot be equal to 0, {\sl i.e.}, $I^{*}>0$ for $\sigma_{pe}^{2}>0$. In order to analyse the case further, we consider the perfect CSIT case, so that $\sigma_{pe}=0$. Recall the definition for $\mathcal{N}\left(\wh{\mathbf{M}}^{H}\right)\backslash \mathbf{0}$ given in the proof of Proposition 2.
Note that such set is spanned by the columns $\mathbf{v}_{j}$ of the unitary matrix $\mathbf{V}$ for $j \in \left\lbrace n_{m}+1, \ldots, N_{t} \right\rbrace $. 
Recall the constraint $\mathrm{rank}(\mathbf{S}^{*})=1$ in the problem \eqref{eqn:problem_analysis_sdp} and assume there is a solution $\mathbf{f}^{*}$ such that $\mathbf{S}^{*}=\mathbf{f}^{*}\left(\mathbf{f}^{*} \right)^{H} $.
Then, the solution $\mathbf{f}^{*}$ should satisfy $\mathbf{f}^{*} \in  \mathcal{N}\left(\wh{\mathbf{M}}^{H}\right)\backslash \mathbf{0}$, $\left(\mathbf{f}^{*} \right)^{H}\mathbf{R}\mathbf{f}^{*}>\rho\frac{\bar{P}_{t}}{N_{p}L}\sigma^{max}$ and $ \left(\mathbf{f}^{*} \right)^{H}\mathbf{f}^{*}=||\mathbf{f}^{*}||^{2}<\frac{\bar{P}_{t}}{N_{p}L}$. Using the inequalities, we write 
\begin{align}
	\rho <  \frac{N_{p}L }{\bar{P}_{t}\sigma^{max}}\left(\mathbf{f}^{*} \right)^{H}\mathbf{R}\mathbf{f}^{*}< \frac{1}{\sigma^{max}}(\tilde{\mathbf{f}}^{*} )^{H}\mathbf{R}\tilde{\mathbf{f}}^{*}, 
	\label{eqn:rho_1} 
\end{align}  
where \mbox{$\tilde{\mathbf{f}}^{*}=\mathbf{f}^{*}/||\mathbf{f}^{*}||$}. It is known that for the positive semi-definite matrix $\mathbf{R}$ and $||\tilde{\mathbf{f}}^{*}||=1$, \mbox{$\left(\tilde{\mathbf{f}}^{*} \right)^{H}\mathbf{R}\tilde{\mathbf{f}}^{*}\leq \sigma^{max}$}, so that the inequality \eqref{eqn:rho_1} does not violate the condition $\rho < 1$.

\textit{Lemma 1:} For any vector $\mathbf{a} \in \mathbb{C}^{N_{t}}$ with elements $a_{i}=0$, $\forall i=\left\lbrace1,2,\ldots,n_{m} \right\rbrace $ and $||a||^{2}=1$, $I^{*}=0$ is a solution for the problem \eqref{eqn:problem_analysis_sdp} if $\sigma_{pe}=0$ and 
\begin{align}
	\rho <  \frac{1}{\sigma^{max}}\sum_{i=1}^{n_{g}}\sigma^{i}\sum_{j=n_{m}+1}^{N_{t}}||a_{j}\left( \mathbf{u}^{i}\right) ^{H} \mathbf{v}^{j}||^{2}.
	\label{eqn:rho_2}
\end{align}  	

\textit{Proof:}  Please see Appendix~\ref{sec:append_lemma1}.  \hspace{2.93cm}$\blacksquare$  

Lemma~1 shows that in order to have a non-empty domain in problem \eqref{eqn:problem2_final_nonconvex}, $I^{thr}$ should be greater than $0$ unless we consider perfect CSIT and \eqref{eqn:rho_2} is satisfied for given $\rho$, $\mathbf{R}$ and $\wh{\mathbf{M}}$ and some $\mathbf{a}$. However, determining such vector is impractical unless $n_{m}=N_{t}-1$, {\sl i.e.,} $\mathrm{rank}\left(\wh{\mathbf{M}}\right)=N_{t}-1$.
\begin{figure}[t!]
	\removelatexerror
	\begin{algorithm}[H]
		\caption{\mbox{$I^{thr}_{l,m,n}\hspace{-0.1cm}=\hspace{-0.1cm}\psi(\rho, \bar{P}_{t},N_{p,l},L,\mathbf{R}_{l,n},\mathbf{\Phi}_{m,n},\mu, \sigma_{pe})$}} 
		\label{alg:algorithm_thr}
		$\mathbf{U}\Sigma \mathbf{U}^{H} \gets \mathrm{SVD}\left(\mathbf{R}_{l,n}\right) $\\
		$\mathbf{V}\Lambda \mathbf{V}^{H} \gets \mathrm{SVD}\left(\mathbf{\Phi}_{m,n}\right) $\\
		$\sigma^{max}=\max_{i}\left\lbrace\sigma^{i}\right\rbrace $\\
		$I^{thr}_{l,m,n}=\rho \frac{\bar{P}_{t}}{N_{p,l}L}\sum_{i=1}^{n_{m}}\lambda^{i}||\left( \mathbf{u}^{max}\right)^{H}\mathbf{v}^{i}||^{2}$  \\
		\If{$\sigma_{pe}=0$}{
			\For{$j=n_{m}+1$ $\mathrm{to}$ $N_{t}$}{
				\If{$\rho < \frac{1}{\sigma^{max}}\sum_{i=1}^{n_{g}}\sigma^{i}||\left( \mathbf{u}^{i}\right) ^{H} \mathbf{v}^{j}||^{2}$}{
					$I^{thr}_{l,m,n}=\mu$ \\
					\textbf{break}
				}        
			}
		}
		\Return $I^{thr}_{l,m,n}$
	\end{algorithm}
	\vspace{-0.5cm}
\end{figure}

From the analysis above, one can conclude that determining the optimal threshold for the interference power is not a straightforward task and may result in an empty domain for the problem \eqref{eqn:problem2_final_nonconvex} when certain conditions are not met. However, a {\sl trivial} threshold that guarantees a non-empty domain can be obtained as follows. We write an upper bound for $\mathrm{tr}\left(\mathbf{S}^{*} \mathbf{\Phi} \right)$ as \cite[Chapter~3]{bhatia_1997}
\begin{align}
	\mathrm{tr}\left(\mathbf{S}^{*} \wh{\mathbf{M}}\wh{\mathbf{M}}^{H}\right)\leq \sum_{i=1}^{N_{t}}\tau^{i}\lambda^{i},
	\label{eqn:thr_upper}
\end{align}
where \mbox{$\tau^{1}\geq\tau^{2}\geq\ldots\geq\tau^{N_{t}}$} are the non-negative eigenvalues of $\mathbf{S}^{*}$. 
Recall the constraint \mbox{$\mathrm{tr}\left(\mathbf{S}^{*} \right)\leq\frac{\bar{P}_{t}}{N_{p}L}$} in the problem formulation \eqref{eqn:problem_analysis_sdp}.
Substituting in \eqref{eqn:thr_upper}, we get 
\begin{align}
	I^{*}\hspace{-0.05cm}\leq\hspace{-0.05cm} \sum_{i=1}^{N_{t}}\hspace{-0.1cm}\tau^{i}\lambda^{i}\hspace{-0.05cm}\leq\hspace{-0.05cm}\lambda^{max}\sum_{i=1}^{N_{t}}\hspace{-0.1cm}\tau^{i}\hspace{-0.05cm}\leq\hspace{-0.05cm} \mathrm{tr}\left(\mathbf{S}^{*}\right)\lambda^{max}\hspace{-0.05cm}\leq\hspace{-0.05cm}\frac{\bar{P}_{t}}{N_{p}L}\lambda^{max}.
	\label{eqn:thr_upper_final}
\end{align}
Note that we have not considered the jamming power constraint \eqref{eqn:jamm_sdp} while determining the upper bound \eqref{eqn:thr_upper_final}. Therefore, the inequality \eqref{eqn:thr_upper_final} is guaranteed to be satisfied by a solution that also satisfies the jamming power constraint, resulting in a non-empty domain for the problem. However,  choosing the threshold in such a trivial fashion to limit the interference to a PU is equivalent to limiting it by the transmit power constraint, thus defying its purpose.  

We propose a threshold calculation method, $I^{thr}_{l,m,n}=\psi(\rho, \bar{P}_{t}, N_{p,l}, L, \mathbf{R}_{l,n}, \mathbf{\Phi}_{m,n},\mu,\sigma_{pe})$ for the pair of AU $l$ and PU $m$, described in Alg.~\ref{alg:algorithm_thr}, to set the threshold according to given system parameters that guarantees a non-empty domain for the problem \eqref{eqn:problem1}. The vector $\mathbf{u}^{max}$ in the algorithm denotes the unit norm eigenvector of $\mathbf{R}_{l,n}$ corresponding to the eigenvalue $\sigma^{max}$. The parameter $\mu_{n}\in\mathbb{R}_{0+}$ is a system parameter, which serves to relax the interference parameters and can be set as desired. The eigen decompositions can be obtained by SVD for the positive semi-definite matrices $\mathbf{R}_{l,n}$ and $\mathbf{\Phi}_{m,n}$, as noted above.

\textit{Proposition 3:} The constraints \eqref{eqn:energy_1} and \eqref{eqn:interference_1} form a non-empty domain for \eqref{eqn:problem2_final_nonconvex} if \mbox{$I^{thr}_{m,n}=\sum_{l \in \mathcal{L}}I^{thr}_{l,m,n}$} for \mbox{$I^{thr}_{l,m,n}=\psi(\rho, \bar{P}_{t}, N_{p,l}, L, \mathbf{R}_{l,n}, \mathbf{\Phi}_{m,n},\mu,\sigma_{pe})$}, $\forall m \in \mathcal{M}$ and 
$\forall n \in \mathcal{S}_{p,l}$.

\textit{Proof:} Please see Appendix~\ref{sec:append_prop3}.  \hspace{2.93cm}$\blacksquare$

In the following corollaries, we provide some insights on the threshold values obtained from Alg.~\eqref{alg:algorithm_thr}.

\textit{Corollary 1:} The threshold obtained by Alg.~\ref{alg:algorithm_thr} is non-trivial in the sense that
\begin{align}
	I^{thr}_{l,m,n}=\psi(\rho, \bar{P}_{t}, N_{p,l}, L, \mathbf{R}_{l,n}, \mathbf{\Phi}_{m,n},\mu,\sigma_{pe}) &\leq \frac{\bar{P}_{t}\lambda^{max}}{N_{p,l}L}, \nonumber 
\end{align}
$l \in \mathcal{L}, m \in \mathcal{M}$, with the equality satisfied if $\mathbf{u}^{max}_{l,n}=\mathbf{v}^{max}_{m,n}$ with $\mathbf{v}^{max}_{m,n}$ being the eigenvector of $\mathbf{\Phi}_{m,n}$ corresponding to eigenvalue $\lambda^{max}$ and $\rho=1$.

\textit{Proof:} Please see Appendix~\ref{sec:append_cor1}.  \hspace{2.93cm}$\blacksquare$
	
\textit{Corollary 2:} The proposed threshold calculation method in Proposition~3 returns the optimal threshold values in \eqref{eqn:thr1} given the corresponding necessary conditions are satisfied and \eqref{eqn:thr2} if $\sigma_{pe}=0$ and $\mathrm{rank}\left(\mathbf{\Phi}_{m,n}\right)=N_{t}-1$.

\textit{Proof:} Please see Appendix~\ref{sec:append_cor2}.  \hspace{2.93cm}$\blacksquare$

\section{Simulation Results}
\label{sec:simulation}
We perform simulations to demonstrate the performance of SUs, PUs and AUs using RSMA, SDMA, and NOMA with the optimized precoders. Note that the optimal precoders for SDMA and NOMA can be obtained by slightly modifying the proposed optimization problem formulation. More specifically, the optimal precoders for SDMA can be obtained by turning off the common stream in the optimization problem formulation, while the optimal precoders for NOMA can be obtained by allocating the message of one user to the common stream and the message of the other user to one private stream without any message splitting. 

We consider a scenario with $N_{t}=4$, $K=2$, $L=1$ and $M=1$. We use Cyclic-Prefix (CP)-OFDM waveform with $N=32$ subcarriers and a CP length of $10\mu$s. The number and indexes of the pilot subcarriers of AU to be jammed are selected as $N_{p}=8$ ans $\mathcal{S}_{p}=\left\lbrace1, 5, 9, 13, 17, 21, 25, 29 \right\rbrace $. We define the metric $SNR=\bar{P}_{t}/(N_{0}N)$. The error variances of the channels of the SUs and the PU are modelled as $\sigma_{ie}^{2}=(SNR)^{-\alpha_{i}}$ and $\sigma_{pe}^{2}=(SNR)^{-\alpha_{p}}$, where $\alpha_{i}$ and $\alpha_{p}$ are the CSIT quality scaling factors \cite{Yang_2013, Joudeh_2016}, and are set as $\alpha_{i}=\alpha_{p}=0.6$ in the simulations unless stated otherwise.
We define the private rate of SU-$k$ for an MC waveform as \mbox{$R_{k}=\frac{1}{N}\sum_{n \in \mathcal{S}}\bar{I}_{k,n}$}, $\forall k\in \mathcal{K}$, following the formulation in \cite{Raleigh_1998} for the carrier cooperative case. This serves as an upper bound for the non-cooperative case since carrier cooperative processing is a more general model \cite{Palomar_2003}. Accordingly, the common rate is defined as \mbox{$R_{c}=\frac{1}{N}\sum_{n \in \mathcal{S}}\bar{I}_{c,n}=\frac{1}{N}\sum_{n \in \mathcal{S}}\sum_{k \in \mathcal{K}}\bar{C}_{k,n}$}. The sum-rate for RSMA is calculated as \mbox{$R_{\mathrm{sum}}=R_{c}+\sum_{k \in \mathcal{K}}R_{k} $}.

We investigate the performance under a frequency-selective channel model. We use the Quadriga Channel Generator \cite{jaeckel_2019} to generate channels according to the 3GPP Urban Macro-Cell channel model \cite{3gpp}. The SUs, PU and AU are placed randomly in a circle with a radius of $400$m around the transmitter and their channels have a delay spread of $1200$ns with $23$ clusters, each cluster consisting of $20$ rays. The channel covariance matrix $R_{1,n}$ is obtained by averaging $\mathbf{g}_{1,n}\mathbf{g}_{1,n}^{H}$ over multiple channel realizations. The OFDM subcarrier spacing is set as $60$kHz. 
	
\subsection{Proposed Thresholds and Feasibility}
\begin{figure}[t!]
	\centering
	\begin{subfigure}[t]{.5\textwidth}
		\centerline{\includegraphics[width=3.5in,height=3.5in,keepaspectratio]{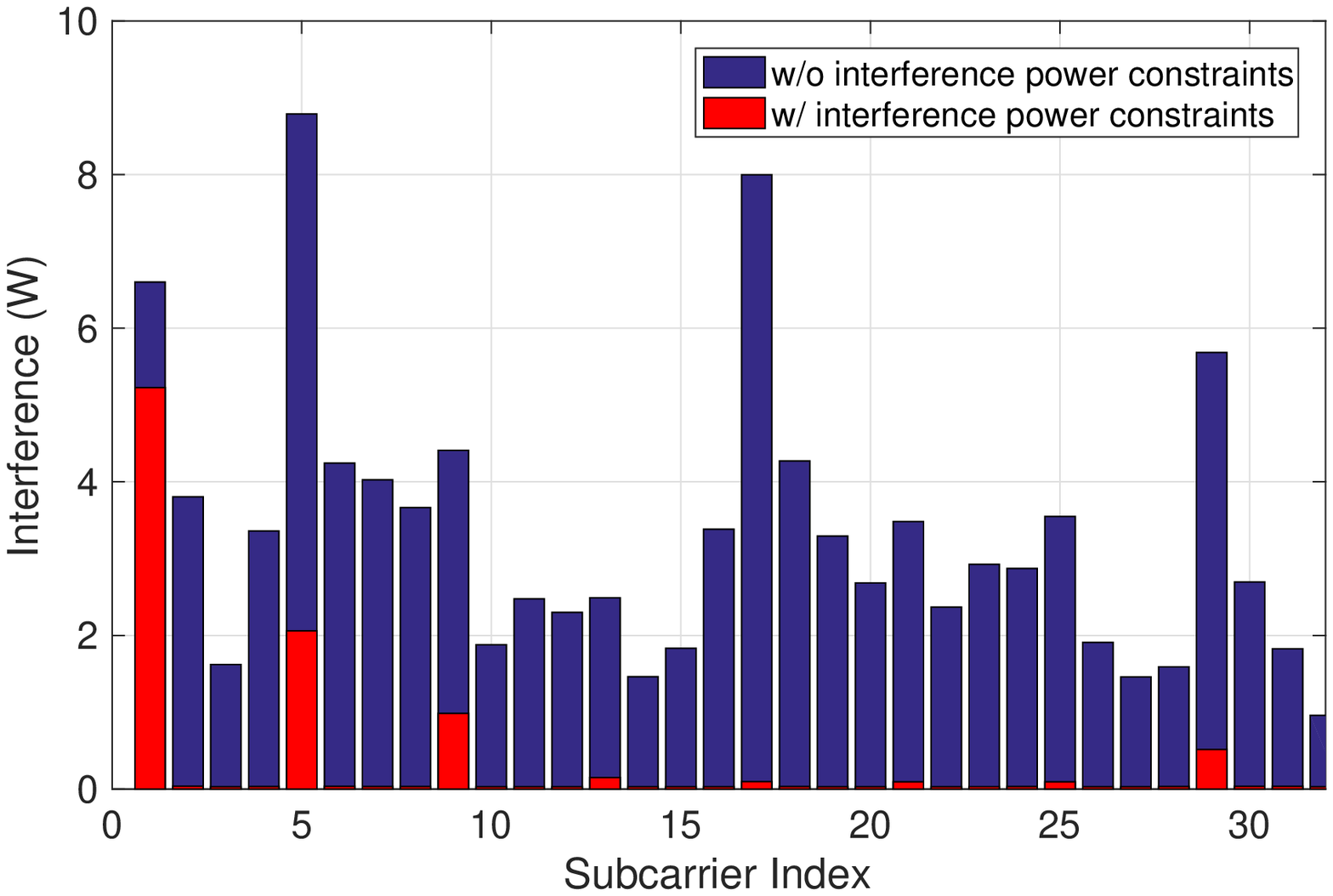}}
		\caption{RSMA.}
		\label{fig:RSMA_Chan1_Int}
	\end{subfigure}
	\begin{subfigure}[t]{.5\textwidth}
		\centerline{\includegraphics[width=3.5in,height=3.5in,keepaspectratio]{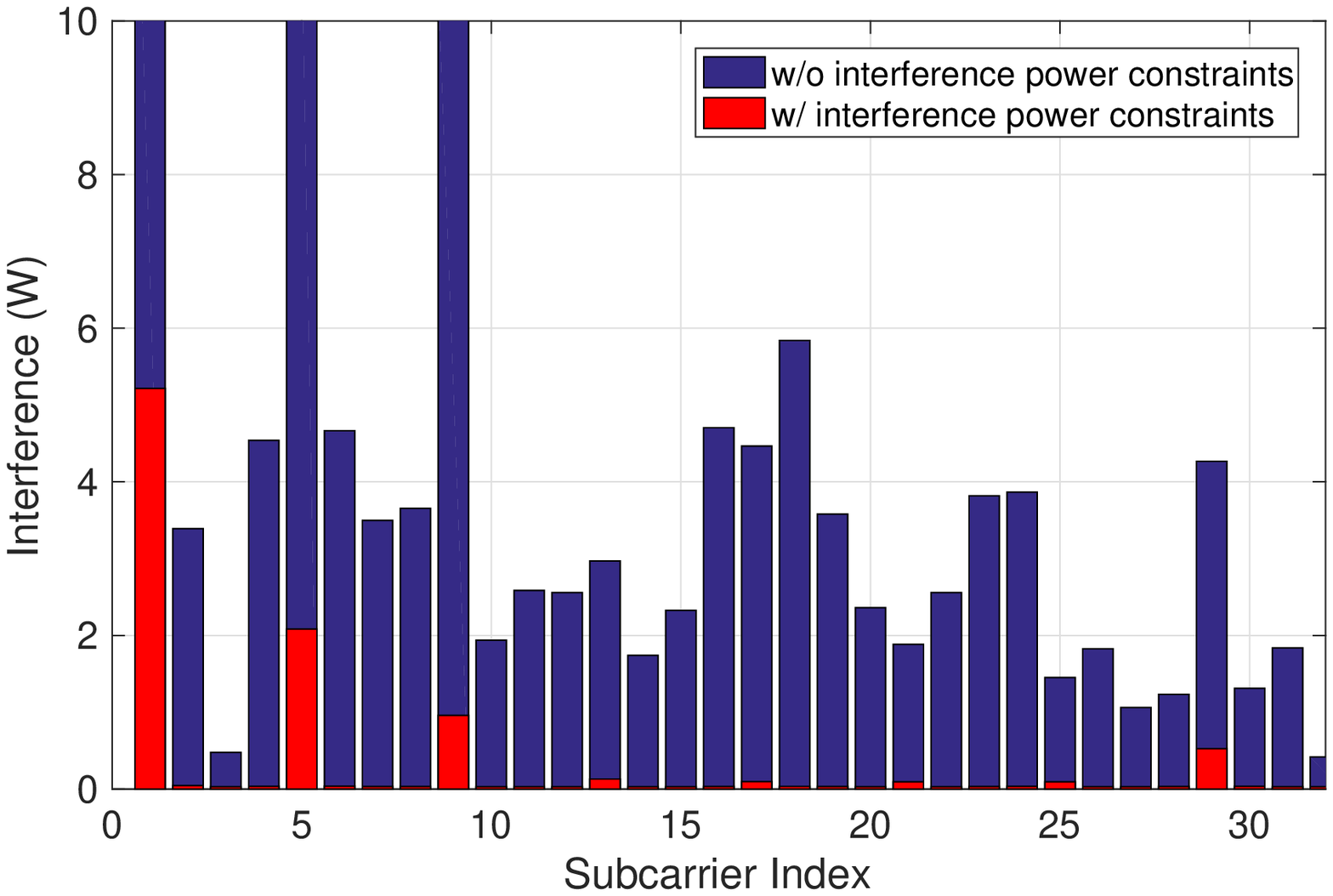}}
		\caption{SDMA.}
		\label{fig:SDMA_Chan1_Int}
	\end{subfigure}
	\begin{subfigure}[t]{.5\textwidth}
		\centerline{\includegraphics[width=3.5in,height=3.5in,keepaspectratio]{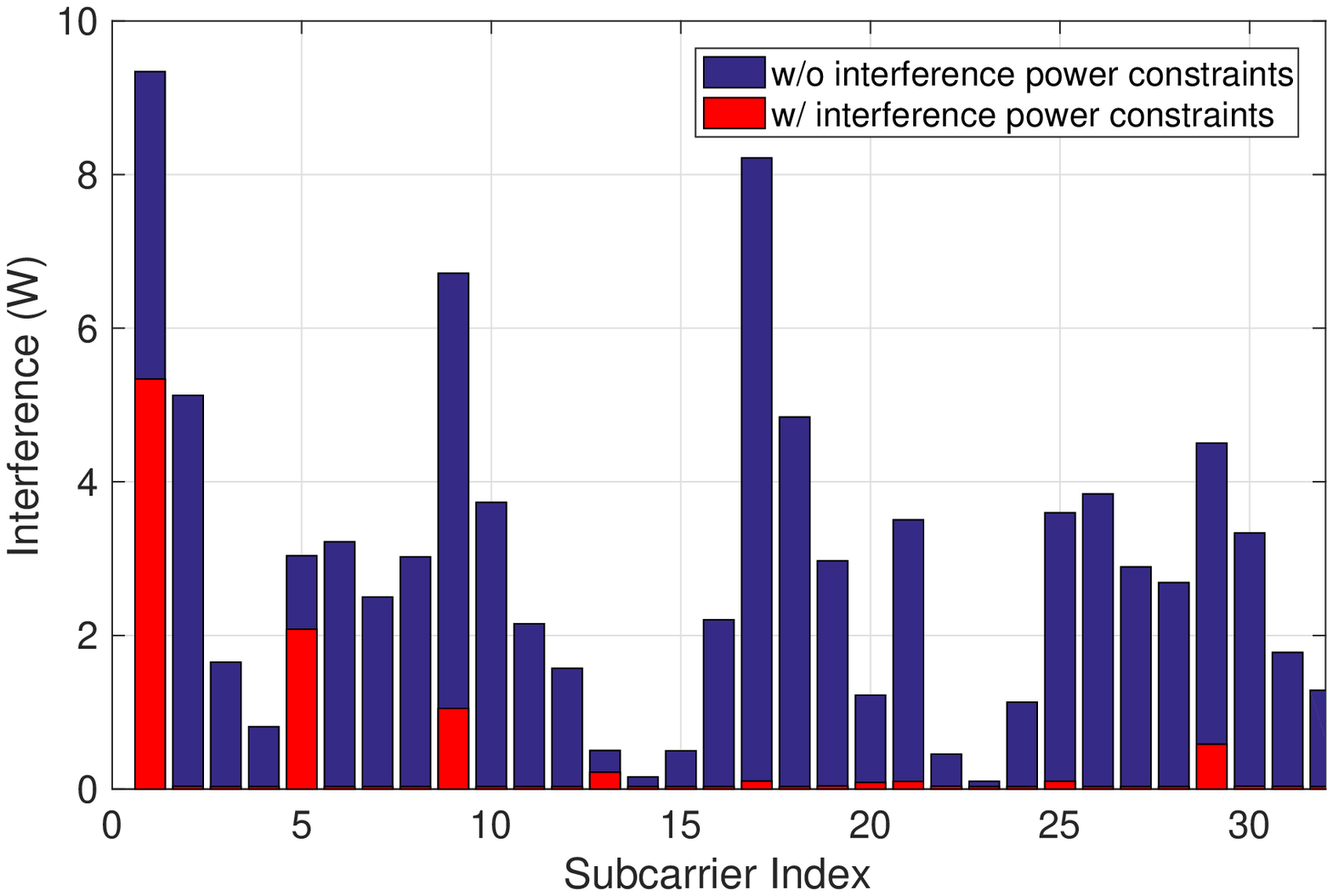}}
		\caption{NOMA.}
		\label{fig:NOMA_Chan1_Int}
	\end{subfigure}
	\caption{Interference power at PU with different schemes.}
	\label{fig:subc_intr}
	\vspace{-0.5cm}
\end{figure}

We start our analysis by verifying that the discussions in Section~\ref{sec:feasibility_analysis} hold (so that we have a feasible problem) and the proposed threshold selection algorithm in Alg.~\ref{alg:algorithm_thr} provides useful threshold values. We investigate the interference power at the PU using precoders obtained with and without interference power constraints, and show that the proposed threshold selection algorithm can obtain threshold values which return a feasible problem and reduce the interference significantly.

Fig.~\ref{fig:subc_intr} shows the interference levels at the subcarriers of PU achieved by RSMA, SDMA, and NOMA under an exemplary random frequency-selective channel realization. We set the parameters $\bar{P}_{t}=100$, $\mu=\bar{P}_{t}/(25N)=0.125$, $\rho=0.45$, and, $R_{th}=0$ bps/Hz.
It is evident from the results that Alg.~\ref{alg:algorithm_thr} returns threshold values that result in a feasible domain for the problem \eqref{eqn:problem1}.  
Furthermore, it is seen from the figures that setting the interference power threshold constraints according to Alg.~\ref{alg:algorithm_thr} reduces the interference to PU significantly, which validates the use of the algorithm.

\subsection{Sum-Rate Performance of SUs}
\begin{figure}[t!]
	\centering
	\centerline{\includegraphics[width=3.5in,height=3.5in,keepaspectratio]{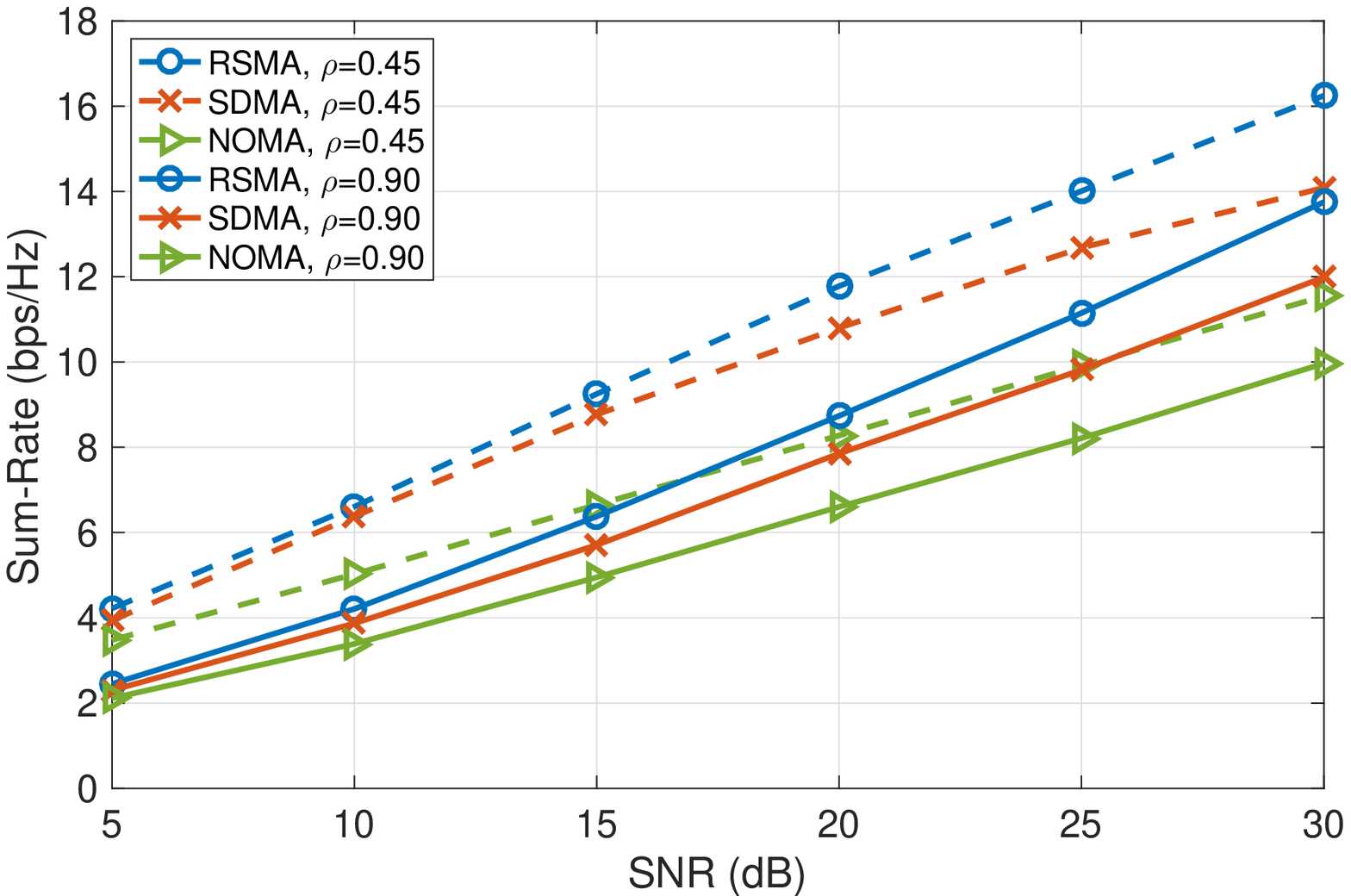}}
	\caption{Sum-rate vs. SNR w/o interference power constraints.}
	\label{fig:Quadriga_ThrCnst_Np8}
	\vspace{-0.5cm}
\end{figure}

\begin{figure}[t!]	
	\centerline{\includegraphics[width=3.5in,height=3.5in,keepaspectratio]{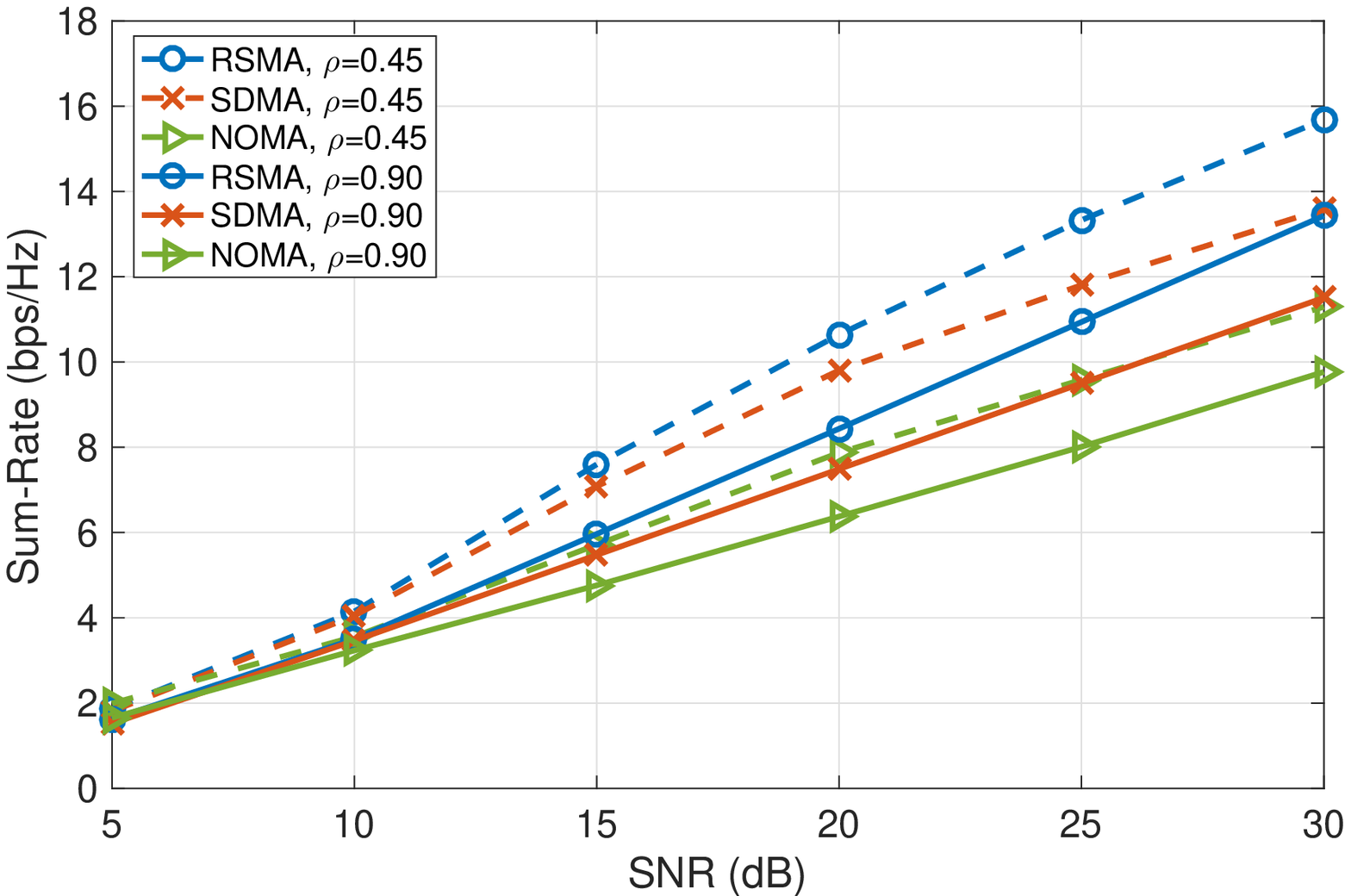}}
	\caption{Sum-rate vs. SNR w/ interference power constraints.}
	\label{fig:Quadriga_ThrCnst_CR_Np8}
	\vspace{-0.5cm}
\end{figure}

With the validation of the feasibility of the problem and the usefulness of the thresholds obtained by Alg.~\ref{alg:algorithm_thr}, 
we move to analyse the sum-rate performance of SUs, and compare the performance of RSMA with SDMA and NOMA. 
Fig.~\ref{fig:Quadriga_ThrCnst_Np8}~and~\ref{fig:Quadriga_ThrCnst_CR_Np8} show the sum-rate performance obtained by RSMA, SDMA, and NOMA for varying $\rho$ values and with or without the interference power constraints, respectively. The noise variance is set as $N_{0}=1/N$ and $R_{th}=0$ bps/Hz in the simulations. The results in the figures are obtained over $100$ realizations of the considered channel model. The interference power constraints are set by Alg.~\ref{alg:algorithm_thr} with $\mu=\bar{P}_{t}/(25N)$. 


The first observation from the figures is the difference in sum-rate for varying values of $\rho$. The jamming power constraint becomes stricter as $\rho$ increases, which results in more transmit power allocated to the pilot subcarriers to be jammed and less power to the data subcarriers. Such power allocation results in a reduced overall sum-rate for all considered schemes.
The second observation from the figures is the performance difference between the schemes. 
RSMA outperforms SDMA and NOMA in all considered cases, owing to its improved interference management capabilities of RSMA framework. Specifically, RSMA achieving a higher throughput without the interference power constraints (Fig.~\ref{fig:Quadriga_ThrCnst_Np8}) show that it can deliver the same jamming power in statistical and imperfect CSIT with higher sum-rate. 
When the interference power constraints are considered (Fig.~\ref{fig:Quadriga_ThrCnst_CR_Np8}), RSMA retains its improved performance, proving that RSMA can perform high performance joint communications and jamming while managing the spatial interference efficiently. 

We note the difference in the performance of the schemes with and without the interference power constraints observed by comparing  Fig.~\ref{fig:Quadriga_ThrCnst_Np8}~and~\ref{fig:Quadriga_ThrCnst_CR_Np8}. Specifically, the performance degradation due to interference power constraints for all considered schemes is significantly higher at low SNR region.
The phenomenon occurs due to the imperfect CSIT of the PU channel, for which the error is higher at low SNR region ($\sigma_{pe}^{2}=(\bar{P}_{t}/N)^{-0.6}$). Observing \eqref{eqn:phi} and \eqref{eqn:int_const}, one can notice that the channel estimation error causes an interference to PU subcarriers that increases with the transmit power on that subcarrier and the estimation error variance (see the second term in \eqref{eqn:phi}). 
Therefore, the sum-rate degradation due to interference power constraints is more severe in the low SNR region than the one in the high SNR region. 

\begin{figure}[t!]
	\centering
	\begin{subfigure}[t]{.24\textwidth}
		\centerline{\includegraphics[width=1.92in,height=1.92in,keepaspectratio]{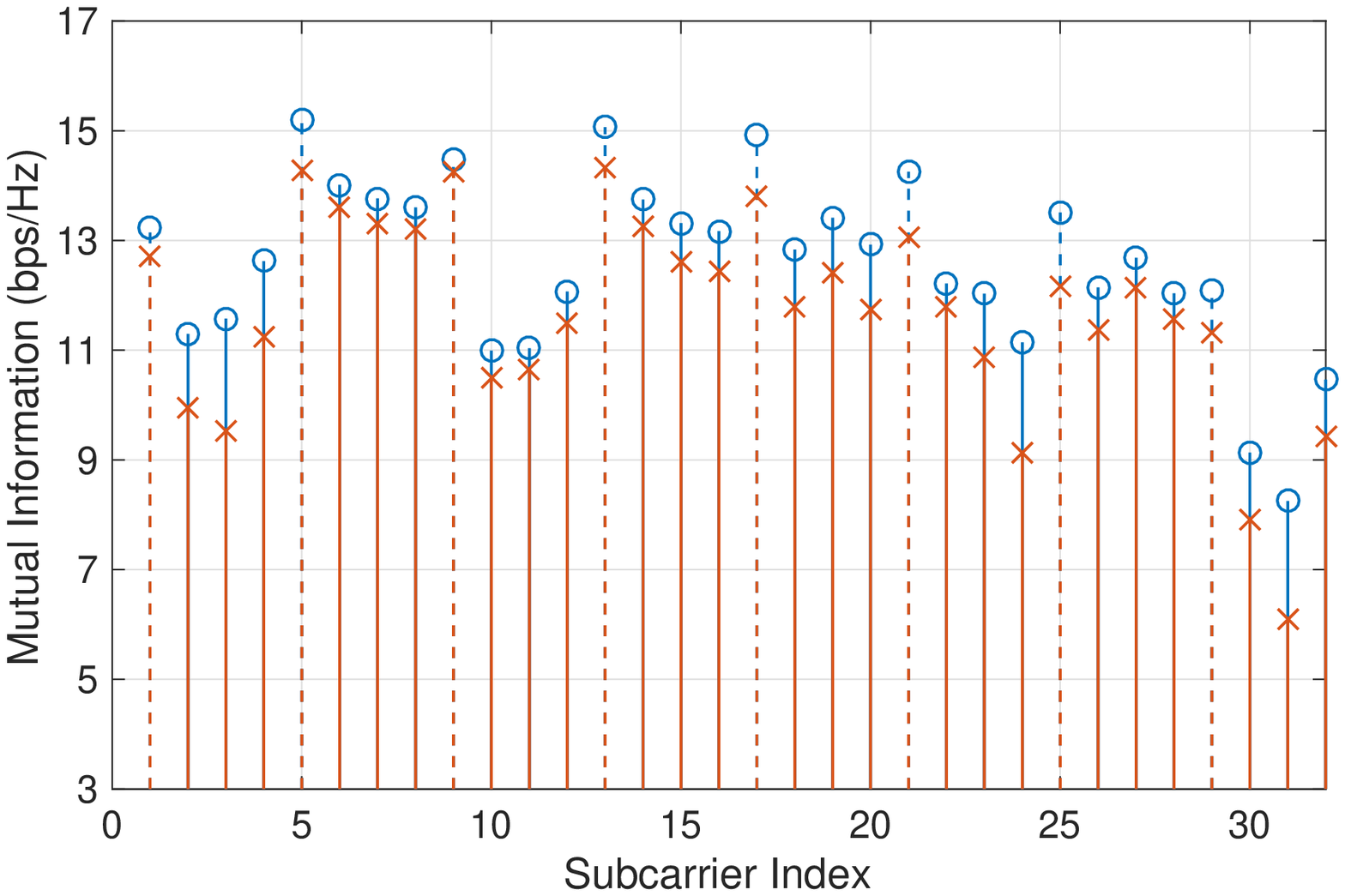}}
		\caption{W/o interference constraints, $\rho=0.45$}
		\label{fig:nocr_rho45}
	\end{subfigure}
	\begin{subfigure}[t]{.24\textwidth}
		\centerline{\includegraphics[width=1.92in,height=1.92in,keepaspectratio]{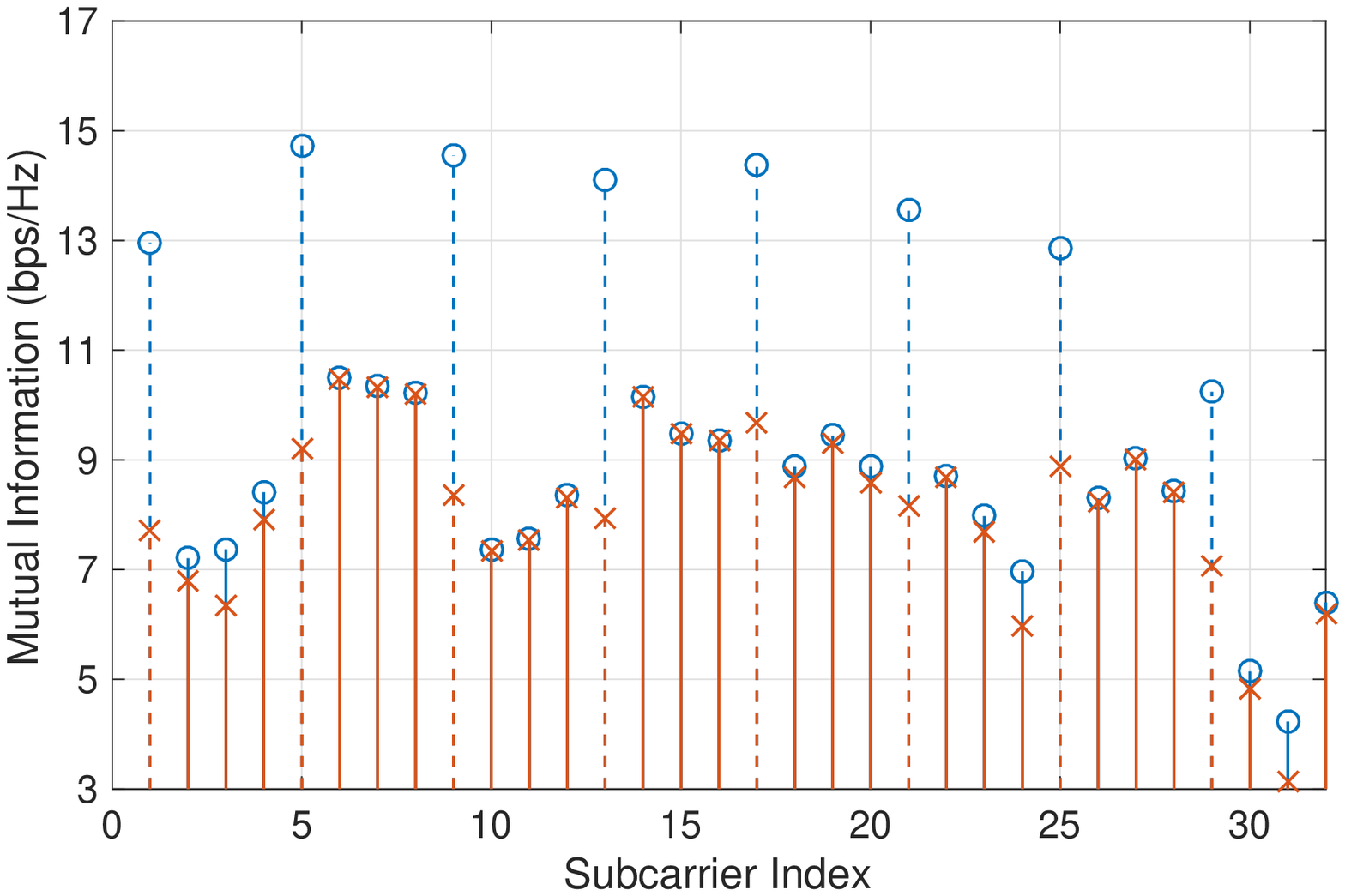}}
		\caption{W/o interference constraints, $\rho=0.90$}
		\label{fig:nocr_rho90}
	\end{subfigure}
	\newline
	\begin{subfigure}[t]{.24\textwidth}
		\centerline{\includegraphics[width=1.92in,height=1.92in,keepaspectratio]{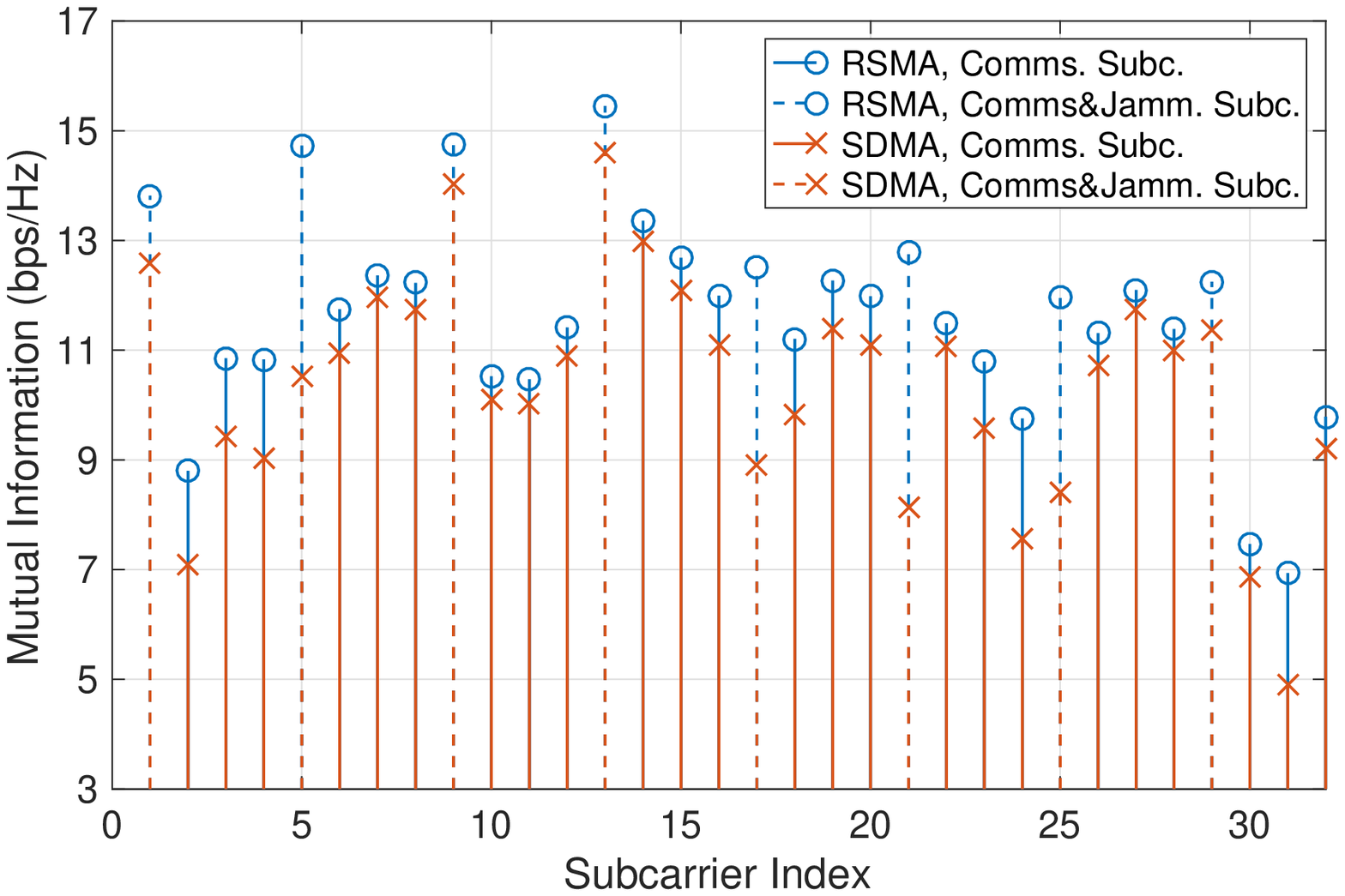}}
		\caption{W/ interference constraints, $\rho=0.45$}
		\label{fig:cr_rho45}
	\end{subfigure}
	\begin{subfigure}[t]{.24\textwidth}
		\centerline{\includegraphics[width=1.92in,height=1.92in,keepaspectratio]{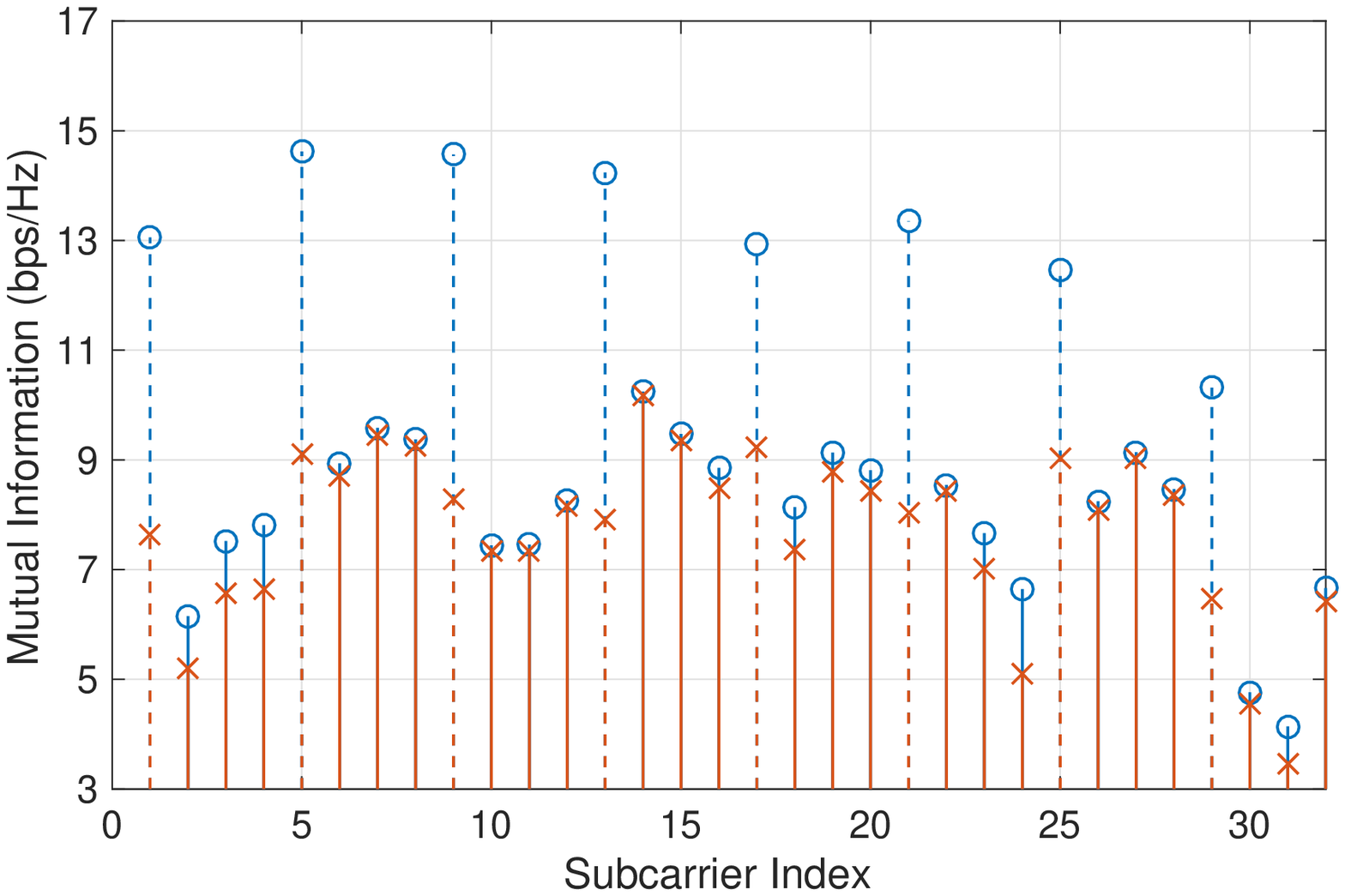}}
		\caption{W/ interference constraints, $\rho=0.90$}
		\label{fig:cr_rho90}
	\end{subfigure}
	\caption{Mutual information at each subcarrier, RSMA vs. SDMA.}
	\label{fig:rates}
	\vspace{-0.5cm}
\end{figure}

\begin{figure}[t!]
	\centering
	\begin{subfigure}[t]{.24\textwidth}
		\centerline{\includegraphics[width=1.92in,height=1.92in,keepaspectratio]{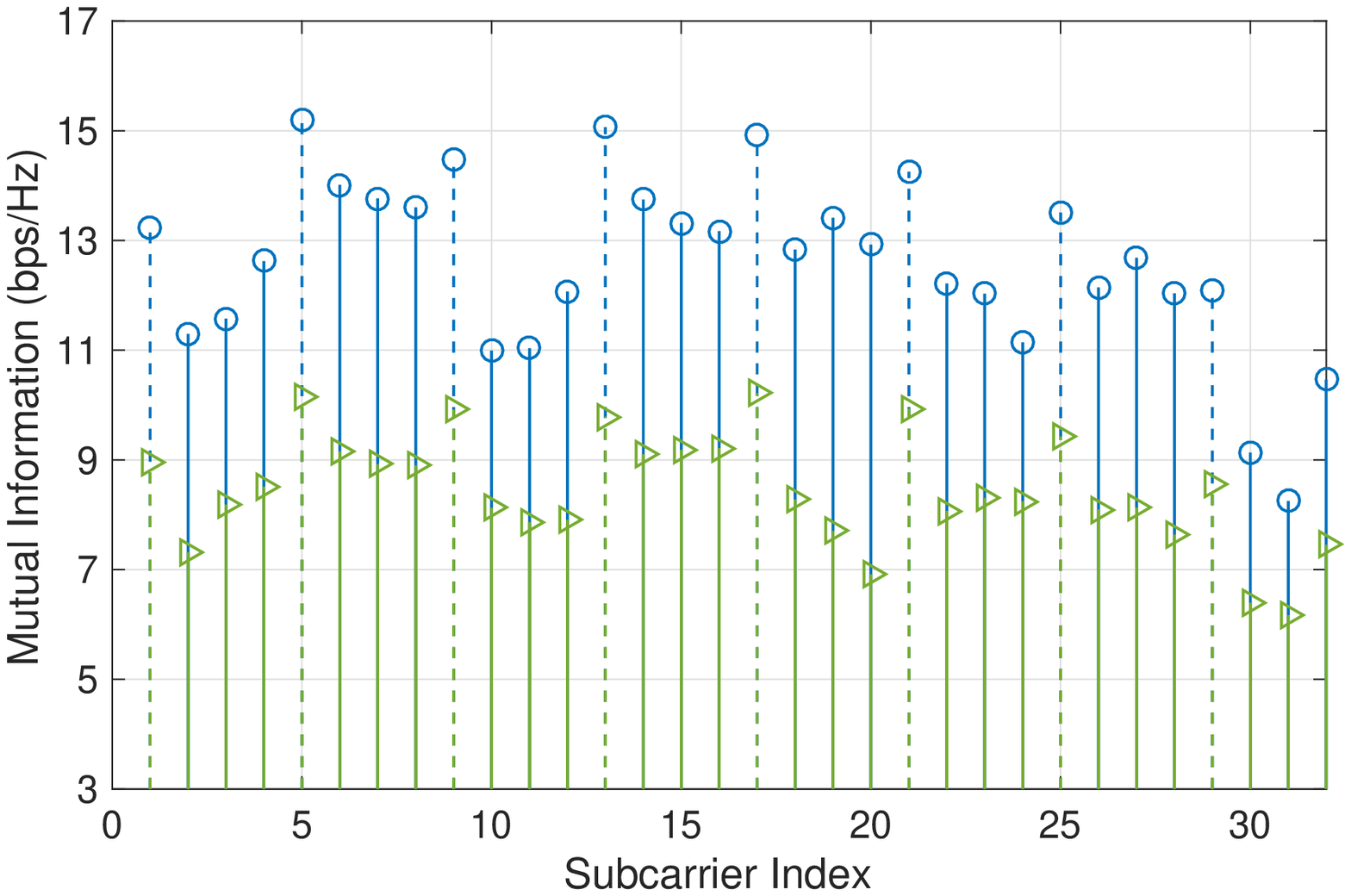}}
		\caption{W/o interference constraints, $\rho=0.45$}
		\label{fig:nocr_rho45_NOMA}
	\end{subfigure}
	\begin{subfigure}[t]{.24\textwidth}
		\centerline{\includegraphics[width=1.92in,height=1.92in,keepaspectratio]{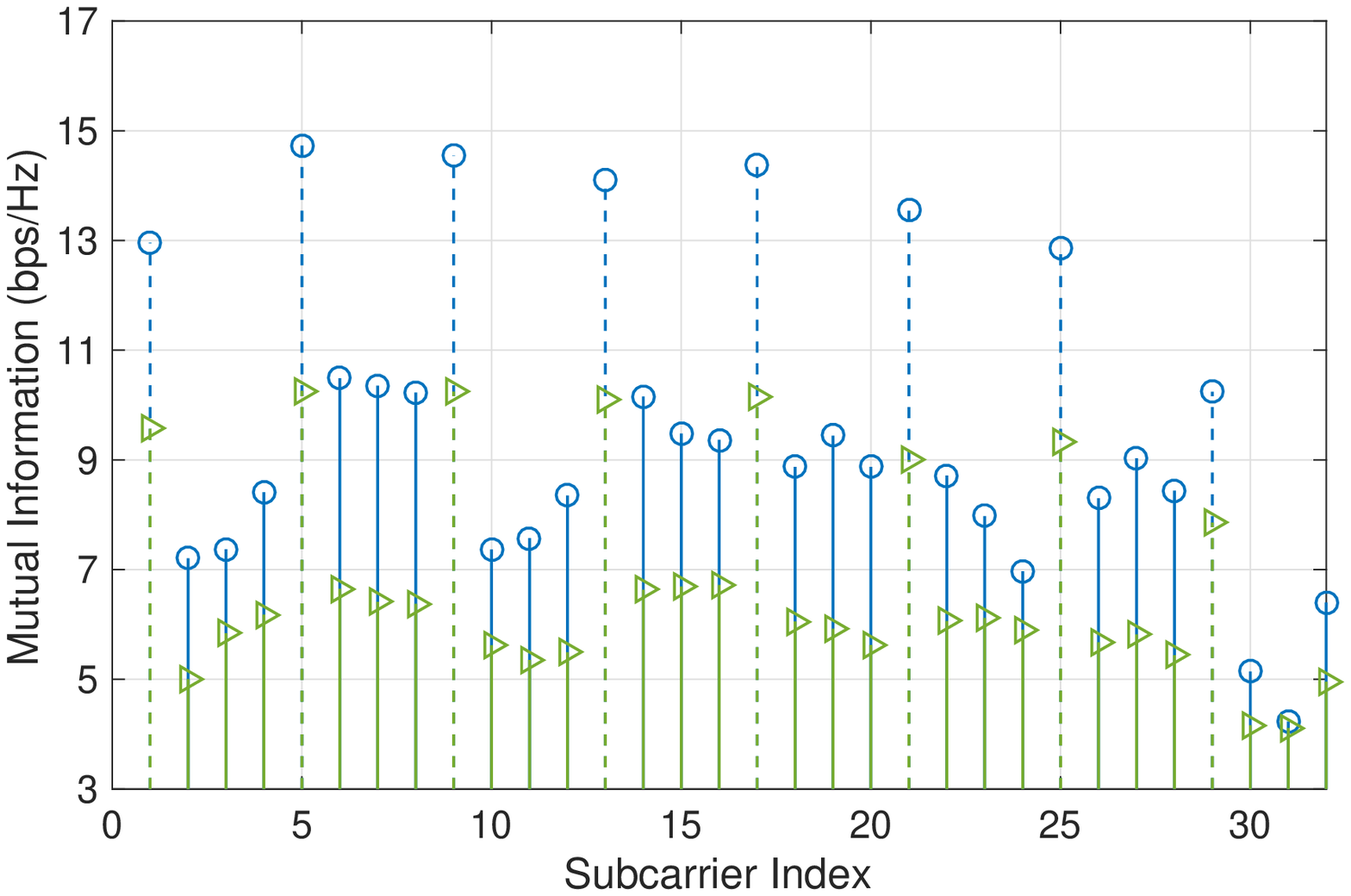}}
		\caption{W/o interference constraints, $\rho=0.90$}
		\label{fig:nocr_rho90_NOMA}
	\end{subfigure}
	\newline
	\begin{subfigure}[t]{.24\textwidth}
		\centerline{\includegraphics[width=1.92in,height=1.92in,keepaspectratio]{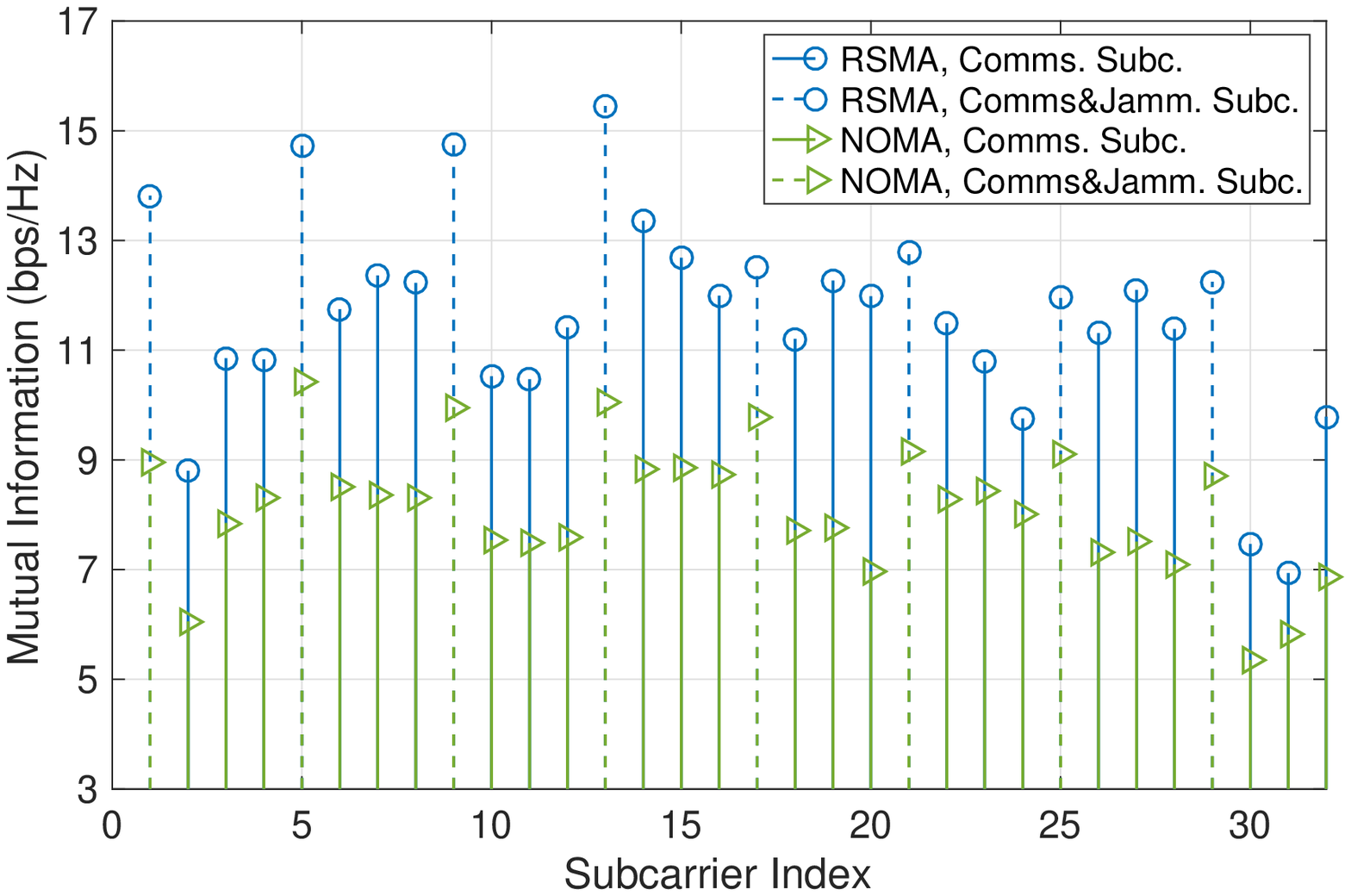}}
		\caption{W/ interference constraints, $\rho=0.45$}
		\label{fig:cr_rho45_NOMA}
	\end{subfigure}
	\begin{subfigure}[t]{.24\textwidth}
		\centerline{\includegraphics[width=1.92in,height=1.92in,keepaspectratio]{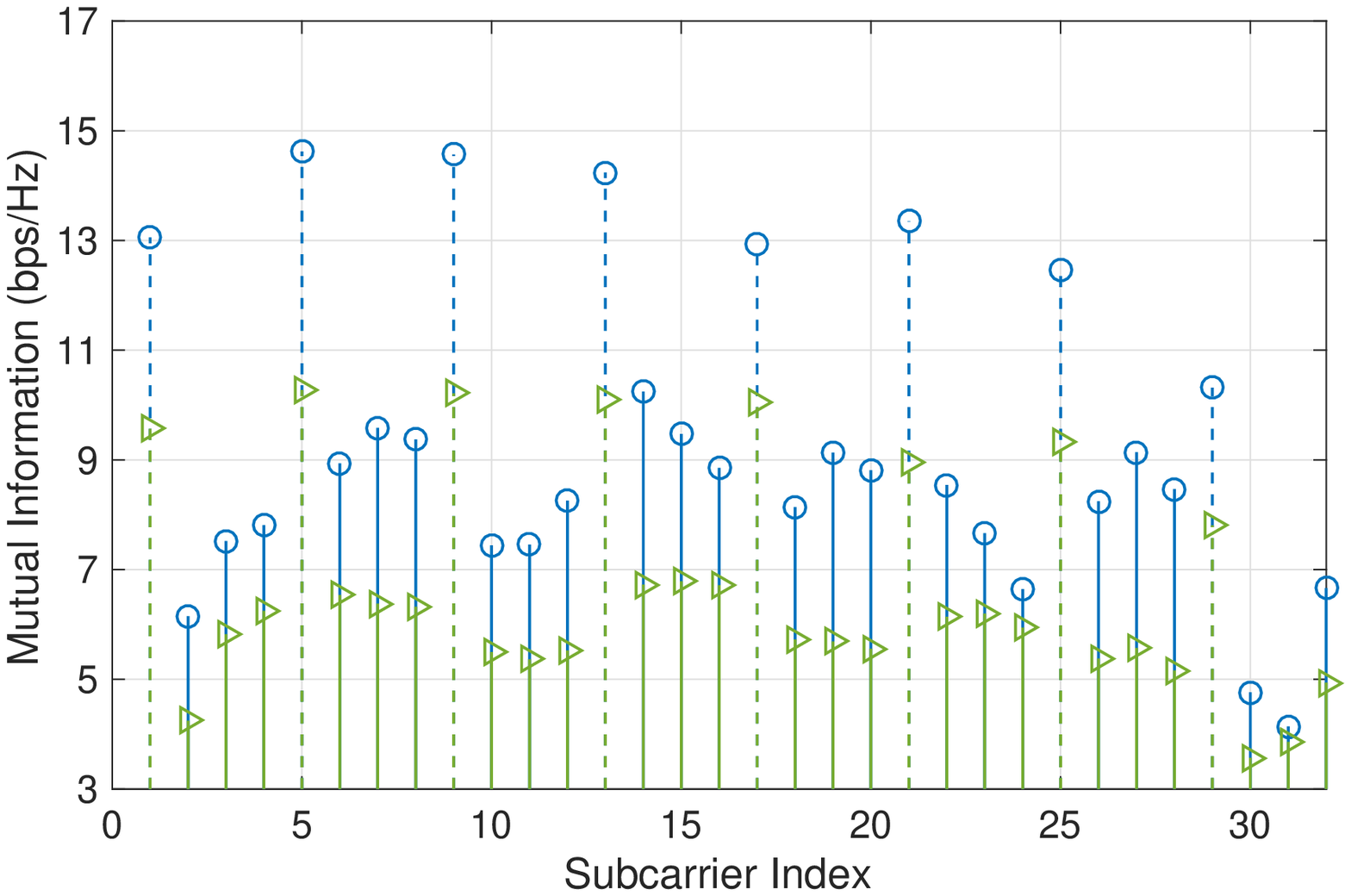}}
		\caption{W/ interference constraints, $\rho=0.90$}
		\label{fig:cr_rho90_NOMA}
	\end{subfigure}
	\caption{Mutual information at each subcarrier, RSMA vs. NOMA.}
	\label{fig:rates_NOMA}
	\vspace{-0.5cm}
\end{figure}

In order to investigate the performance of the schemes in more detail, Fig.~\ref{fig:rates}~and~\ref{fig:rates_NOMA} compare the mutual information at each subcarrier achieved by RSMA, SDMA, and NOMA with and without interference power constraints for $\rho=0.45$ and $\rho=0.90$. We consider the channel realization in Fig.~\ref{fig:SDMA_Chan1_Int}~and~\ref{fig:RSMA_Chan1_Int} with $\bar{P}_{t}=100$ for the analysis. As observed from the figures, RSMA achieves a higher mutual information at each subcarrier than SDMA and NOMA. The difference in the mutual-information achieved by RSMA and the other schemes at pilot subcarriers of AU increases as $\rho$ increases, implying that RSMA can focus power more efficiently under statistical CSIT than SDMA and NOMA. 

The difference in mutual information achieved by RSMA, SDMA, and NOMA on data subcarriers decrease as $\rho$ increases. The reason for such behaviour is the decreasing signal power on data subcarriers with increasing jamming power threshold, resulting in the system operating in the noise-limited region on these subcarriers. As stated above, the mutual information gain by RSMA becomes more significant at pilot subcarriers, where the signal power is increased and the system operates in interference-limited region. 
The mutual information gain of RSMA is preserved when the interference power constraints are considered.

\begin{figure}[t!]	
	\centerline{\includegraphics[width=3.5in,height=3.5in,keepaspectratio]{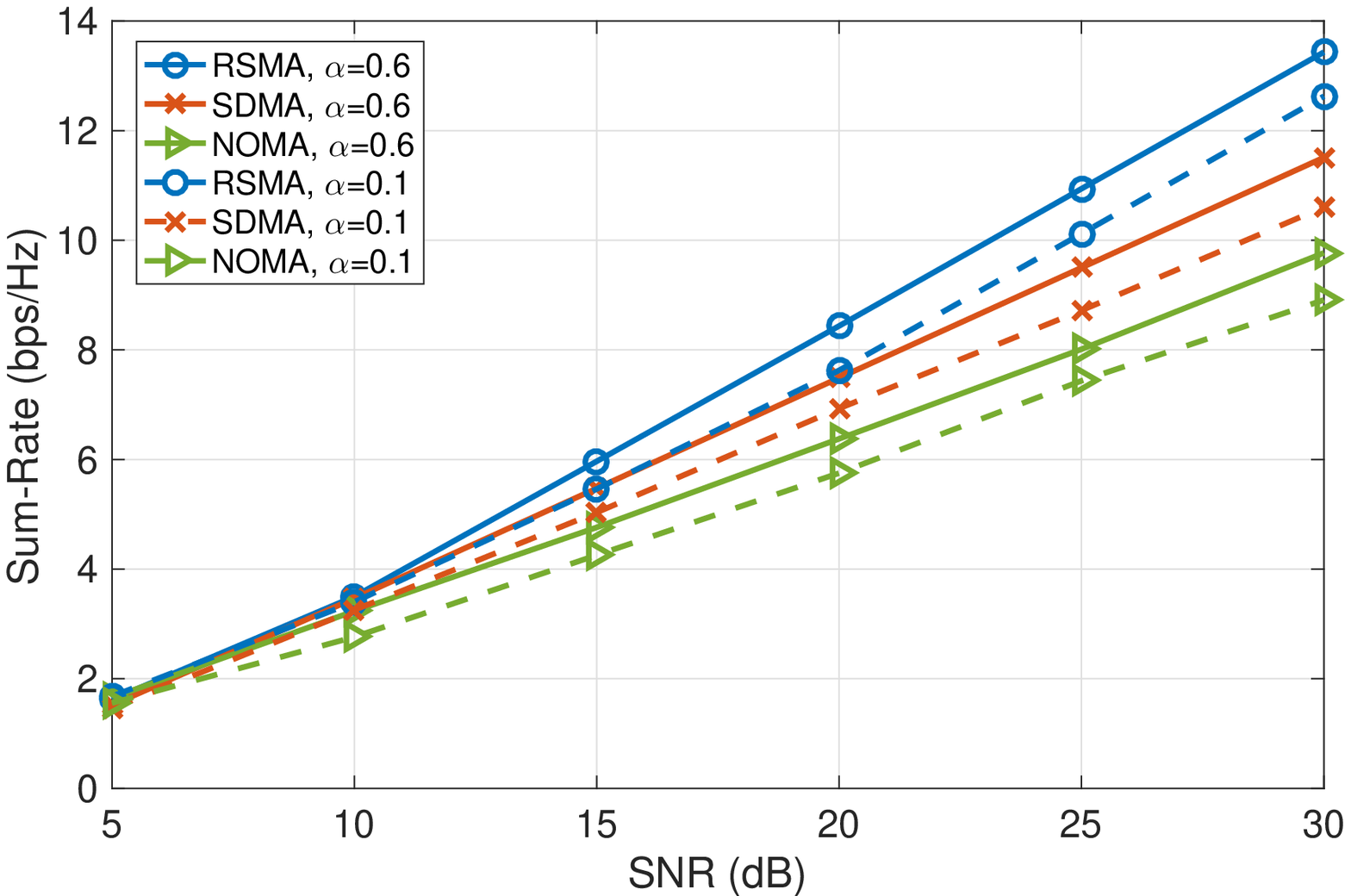}}
	\caption{Sum-rate vs. SNR for different CSIT qualities.}
	\label{fig:Quadriga_ThrCnst_CR_Np8_alpha01}
	\vspace{-0.5cm}
\end{figure}

\begin{figure}[t!]	
	\centerline{\includegraphics[width=3.5in,height=3.5in,keepaspectratio]{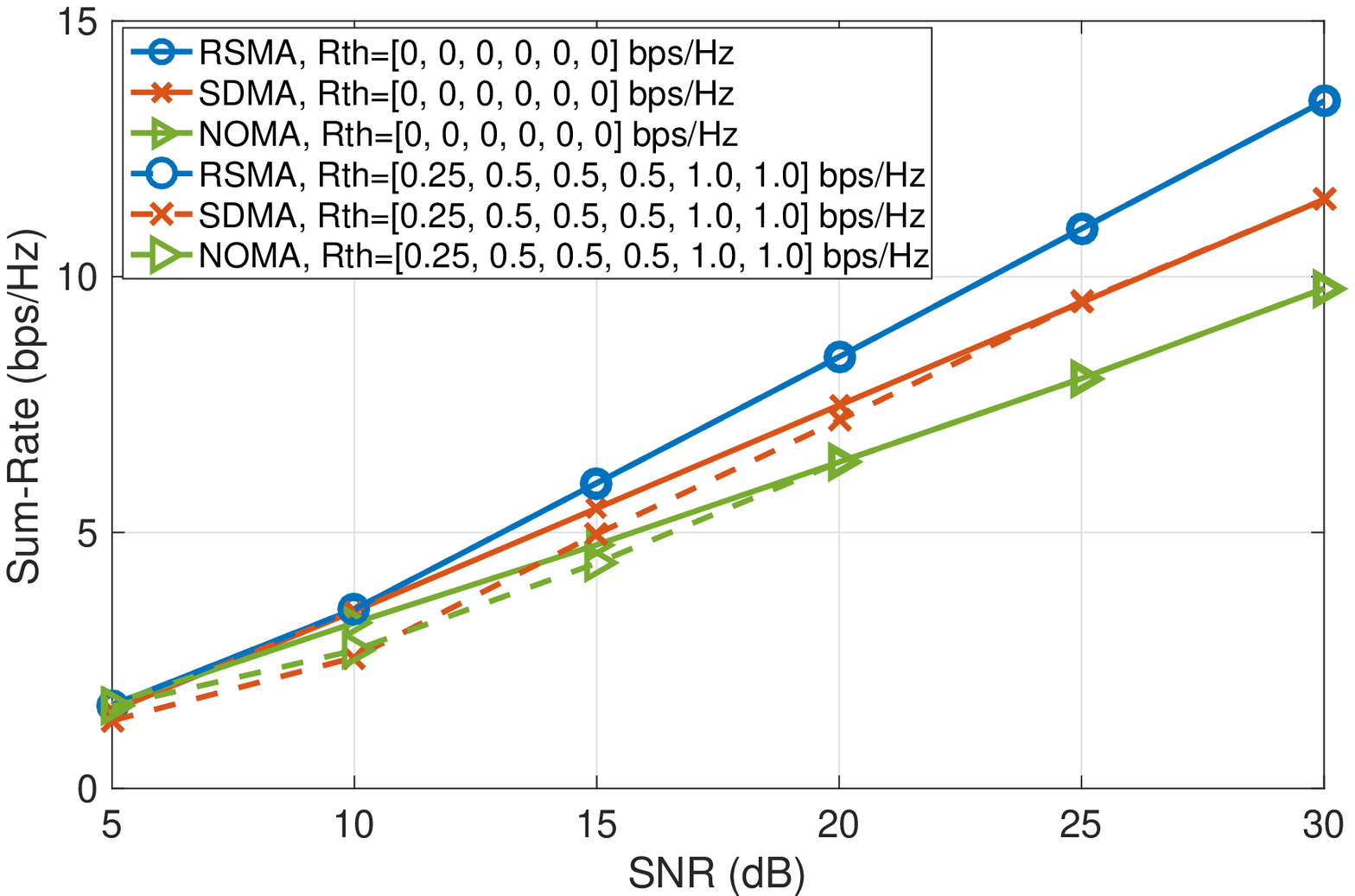}}
	\caption{Sum-rate vs. SNR for different $R_{th}$ values.}
	\label{fig:Quadriga_ThrCnst_CR_Np8_RateCnstr}
	\vspace{-0.5cm}
\end{figure}

We continue to investigate the sum-rate performance of the considered schemes by varying the CSIT quality and QoS constraint threshold. Fig.~\ref{fig:Quadriga_ThrCnst_CR_Np8_alpha01} shows the performance variation of RSMA, SDMA, and NOMA with decreasing CSIT quality ($\alpha=0.1$). As expected, RSMA outperforms the other considered schemes owing to its robust interference management capabilities. Fig.~\ref{fig:Quadriga_ThrCnst_CR_Np8_RateCnstr} gives the performance of the considered schemes with $R_{th}=0.25$ bps/Hz for SNR$=5$dB, $R_{th}=0.5$ bps/Hz for $10$dB $\leq$ SNR $\leq 20$dB, and $R_{th}=1.0$ bps/Hz for SNR $\geq 25$dB. One can observe from the results that the performance of RSMA is not affected by the considered QoS threshold, implying that it can support robust multiple access without any loss in sum-rate. On the other hand, SDMA and NOMA suffer from performance loss, especially at low and medium SNR regimes. 

\subsection{Performance Comparison with Benchmark Schemes}
In this section, we compare the performance of SUs, AU and PU with the proposed system model and with specific benchmark schemes in order to demonstrate the effectiveness of the system.

Although the effectiveness of pilot jamming on MC waveforms is known in the literature \cite{Shahriar_2015, Shahriar_2016, clancy_2011, miller_2012, patel_2004, han_2008}, we devise a simple setup to demonstrate it in our system model. 
For performance comparison, we consider barrage jamming as the benchmark scheme, which performs jamming by focusing power over the whole communication band in a uniform manner.
We implement barrage jamming in our system by assigning $\mathcal{S}_{p}=\mathcal{S}$ and calculating the thresholds $J^{thr}_{n}$ and $I^{thr}_{n}$, $\forall n \in \mathcal{S}_{p}$, accordingly.
We investigate the sum-rate performance for the SUs and uncoded BER performance for the AU and PU. 

Assume that the AU is communicating with a single-antenna user in its corresponding network using an MC waveform over a flat-fading channel. The transmitted symbols are chosen from the QPSK modulation alphabet with unit power and equal power allocation is performed over all subcarriers. The received signal at subcarrier-$n$ at the AU is expressed as
\begin{align}
		y_{AU,n}=\sqrt{E_{s}/N}h_{AU}x_{AU,n}+z_{AU,n}+\eta_{AU,n}, \ \forall n \in \mathcal{S}, \nonumber 
\end{align}
\hspace{-0.1cm}where $E_{s}$ is the total received symbol energy at the AU, $h_{AU}$ is the frequency-response of the AU-AU channel at each subcarrier, $z_{AU,n} \sim \mathcal{CN}(0,N_{0})$ is the AWGN component and $\eta_{AU,n}=\mathbf{g}_{n}^{H}\mathbf{p}_{c,n}+\sum_{k \in \mathcal{K}}\mathbf{g}_{n}^{H}\mathbf{p}_{k,n}+\sum_{l^{\prime} \in \mathcal{L}}\mathbf{g}_{n}^{H}\mathbf{f}_{l^{\prime},n}$ is the jamming power focused by the secondary transmitter. We consider that the AU performs channel estimation over $8$ pilots, so that the estimate of the flat-fading AU-AU channel is given by
\begin{align}
	h_{AU}=\sqrt{1\minus\sigma_{AU}^{2}}\widehat{h}_{AU}+\sigma_{AU}\widetilde{h}_{AU}, \nonumber 
\end{align}
where $\sigma_{AU}=\left(\frac{1}{N_{p}}\sum_{n \in \mathcal{S}_{p}}SINR_{AU,n}\right)^{-0.6}$, $SINR_{AU,n}=\frac{|	h_{AU}|^{2}E_{s}/N}{N_{0}+J_{AU,n}}$ and $J_{AU,n}=|\mathbf{g}_{n}^{H}\mathbf{p}_{c,n}|^{2}\hspace{-0.1cm}+\hspace{-0.1cm}\sum_{k \in \mathcal{K}}\hspace{-0.1cm}|\mathbf{g}_{n}^{H}\mathbf{p}_{k,n}|^{2}\hspace{-0.1cm}+\hspace{-0.1cm}\sum_{l^{\prime} \in \mathcal{L}}\hspace{-0.1cm}|\mathbf{g}_{n}^{H}\mathbf{f}_{l^{\prime},n}|^{2}$. The AU performs detection of the received QPSK symbols by performing zero-forcing equalization using the obtained channel estimate, {\sl i.e.,} $\widehat{y}_{AU,n}=y_{AU,n}/\widehat{h}_{AU}$. 
\begin{figure}[t!]	
	\centerline{\includegraphics[width=3.5in,height=3.5in,keepaspectratio]{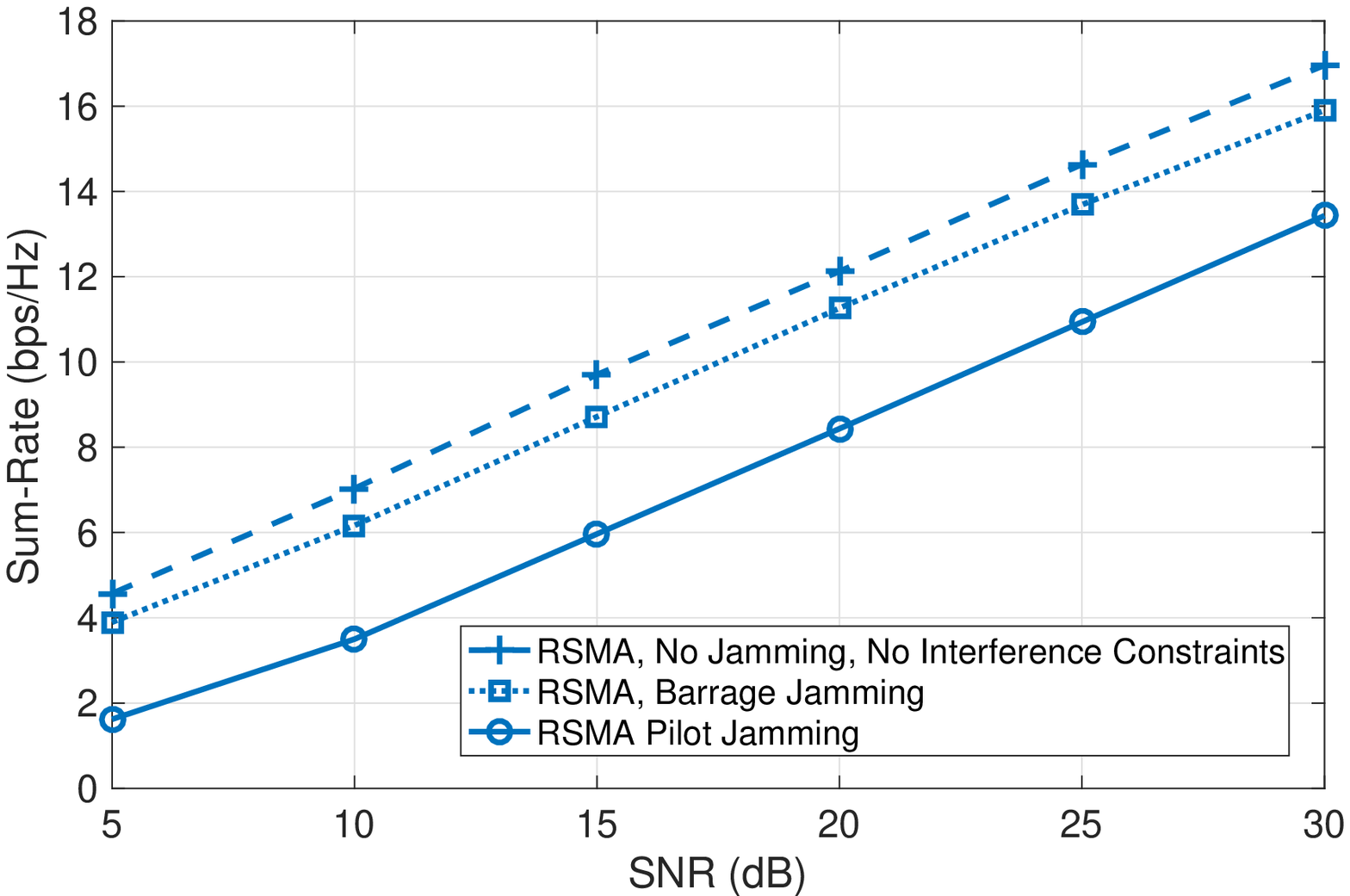}}
	\caption{Sum-rate of SUs with different jamming strategies.}
	\label{fig:Quadriga_ThrCnst_CR_Np8_diffjamm}
	\vspace{-0.5cm}
\end{figure}

We consider the scenario where the secondary transmitter performs transmission with $P_{t}=5$ dBW and $\rho=0.9$ using RSMA. Fig.\ref{fig:Quadriga_ThrCnst_CR_Np8_diffjamm} gives the sum-rate performance comparison for RSMA with pilot jamming and barrage jamming. The figure also gives the performance of conventional RSMA (without jamming and interference power constraints) with an MC waveform obtained by the AO-WMMSE method in \cite{Joudeh_2016}. As expected, the figure demonstrates that performing joint communications and jamming at the secondary transmitter degrades the sum-rate performance. Furthermore, performing communications with barrage jamming yields a higher sum-rate performance since the transmit energy is distributed in a more uniform manner over the subcarriers compared to the pilot jamming case. 
\begin{figure}[t!]	
	\centerline{\includegraphics[width=3.5in,height=3.5in,keepaspectratio]{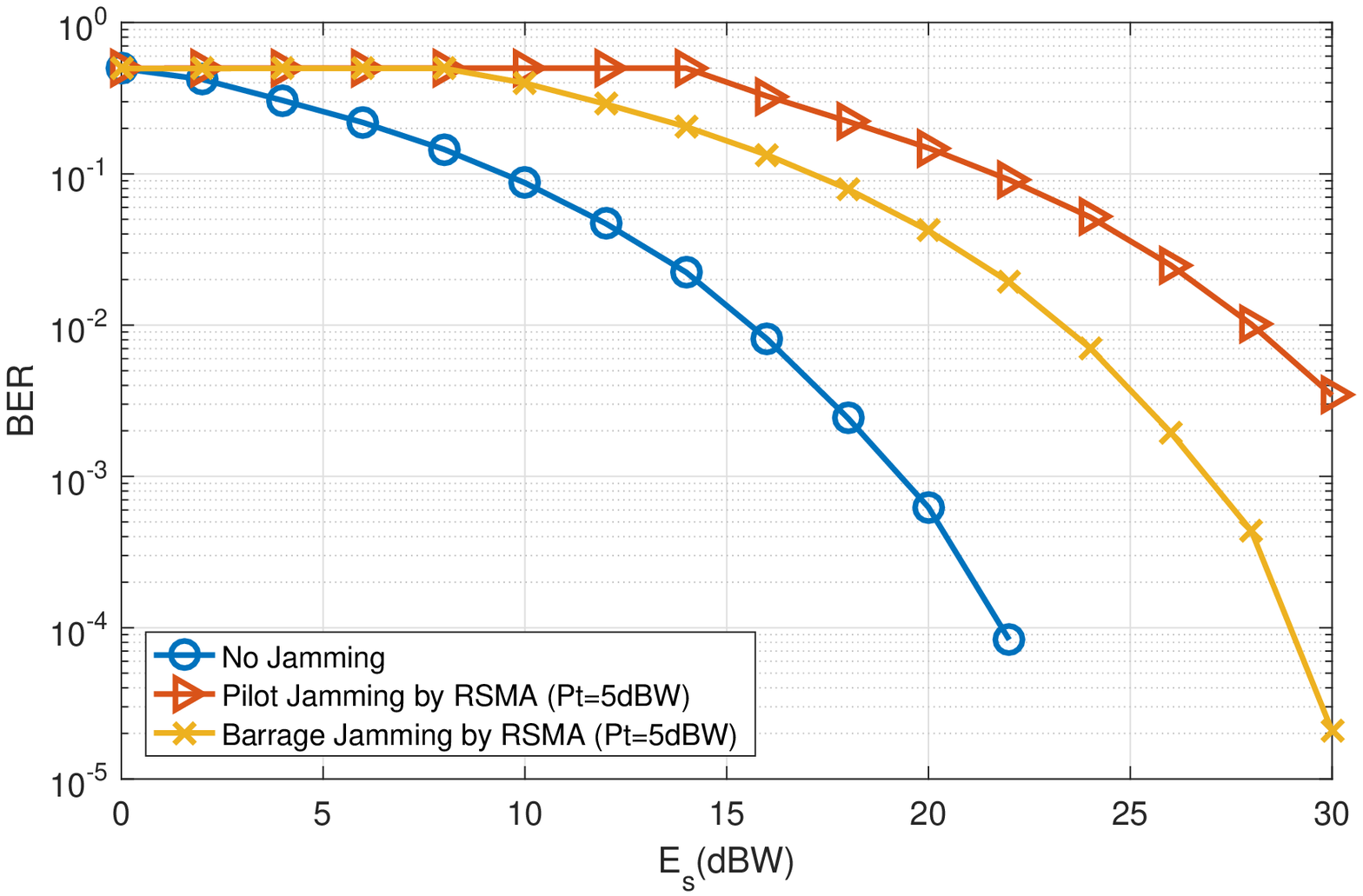}}
	\caption{BER performance of AU with different jamming strategies.}
	\label{fig:AU_BER}
	\vspace{-0.5cm}
\end{figure}

Next, we investigate the uncoded BER performance of AU under no jamming, pilot jamming, and barrage jamming. Fig.\ref{fig:AU_BER} gives the BER results with respect to total symbol energy $E_{s}$ for the considered cases. The case with no jamming demonstrates the performance of the AU-AU communications in the considered flat-fading channel and with the channel estimation error under additive noise. It is observed that pilot jamming affects the performance more severely than barrage jamming, as expected.

\begin{figure}[t!]	
	\centerline{\includegraphics[width=3.5in,height=3.5in,keepaspectratio]{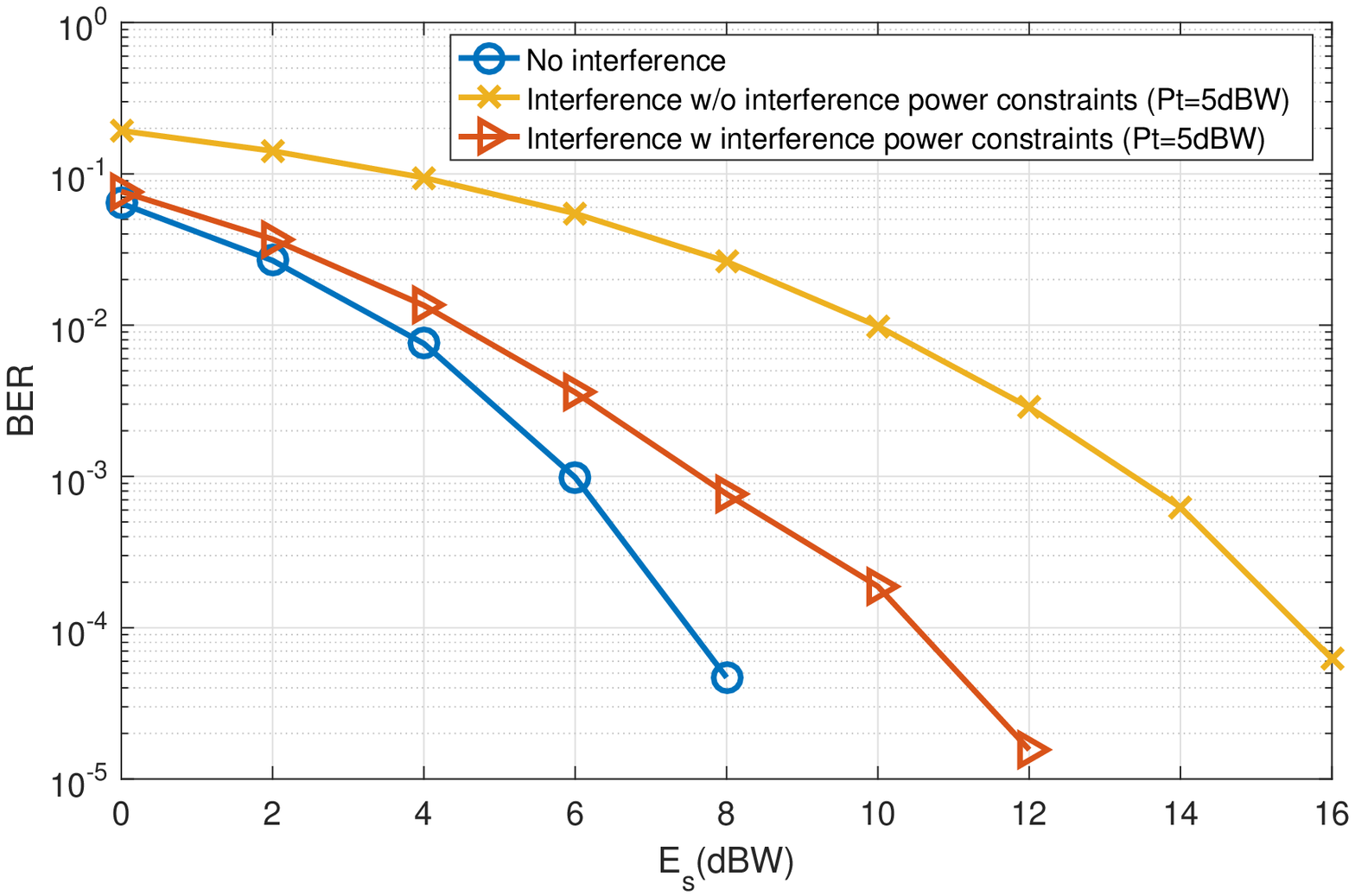}}
	\caption{BER performance of PU with different jamming strategies.}
	\label{fig:PU_BER}
	\vspace{-0.5cm}
\end{figure}

Finally, we move to investigate the performance of PU. It was shown in Fig.~\ref{fig:subc_intr} that the interference power thresholds reduced the interference at PU significantly. Now, we investigate the uncoded BER performance of the PU under no interference and interference due to pilot jamming with and without interference power constraints. Similar to the scenario for AU-AU communications, we assume the AU is communicating with a single-antenna user in its corresponding network using an MC waveform over a flat-fading channel. The transmitted symbols are chosen from the QPSK modulation alphabet with unit power and equal power allocation is performed over all subcarriers. The received signal at subcarrier-$n$ at the AU is expressed as
\begin{align}
		y_{PU,n}=\sqrt{E_{s}/N}h_{PU}x_{PU,n}+z_{PU,n}+\eta_{PU,n}, \ \forall n \in \mathcal{S}, \nonumber 
\end{align}
\hspace{-0.1cm}where $E_{s}$ is the total received symbol energy at the PU, $h_{PU}$ is the frequency-response of the PU-PU channel at each subcarrier, $z_{PU,n} \sim \mathcal{CN}(0,N_{0})$ is the AWGN component and $\eta_{PU,n}=\mathbf{m}_{n}^{H}\mathbf{p}_{c,n}+\sum_{k \in \mathcal{K}}\mathbf{m}_{n}^{H}\mathbf{p}_{k,n}+\sum_{l^{\prime} \in \mathcal{L}}\mathbf{m}_{n}^{H}\mathbf{f}_{l^{\prime},n}$ is the interference due to the transmission of the secondary transmitter. In order to observe the performance degradation due to interference more clearly, we assume that the PU has perfect Channel State Information at Receiver (CSIR), so that, the performance degradation is purely due to the interference over the data subcarriers. The PU performs detection of the received QPSK symbols by performing zero-forcing equalization using the perfect CSIR, {\sl i.e.,} $\widehat{y}_{PU,n}=y_{PU,n}/{h_{PU}}$.
	
We consider the scenario where the secondary transmitter performs transmission with $P_{t}=5$ dBW and $\rho=0.45$ using RSMA. Fig.\ref{fig:PU_BER} gives the BER results with respect to total symbol energy $E_{s}$ for the abovementioned cases. The case with no interference serves as a benchmark for the performance of the PU-PU communications in the considered flat-fading channel. It is observed from the figure that using the interference constraints cause significantly less performance degradation compared to the case without constraints, validating the usefulness of interference power constraints and the proposed algorithm to determine $I^{thr}$. 

\section{Conclusion}
\label{sec:conclusion}
In this work, we consider RSMA for multi-antenna multiple-access joint communications and jamming in an RF-congested CR network using MC waveforms. 
We formulate a mutual information maximization problem with minimum focused jamming power constraint on pilot subcarriers of the AUs and maximum interference power constraint for the PUs, 
under the practical case of imperfect CSIT for the SUs and PUs and statistical CSIT for the AUs. 
We propose an AO-ADMM-based algorithm to deal with the formulated non-convex problem. 
An analysis based on KKT conditions is provided to determine the optimal interference and jamming power constraints to guarantee the feasibility of the problem. 
Based on this analysis, we propose a practical interference threshold calculation algorithm for given jamming power parameters and characteristics of the channels of AUs and PUs. 
We perform simulations to compare the sum-rate performance of RSMA and SDMA under a realistic 3GPP frequency-selective channel model. 
Our simulation results show that RSMA can achieve significantly higher sum-rate than SDMA for joint communications and jamming with MC waveforms in CR networks. 

\section{Appendix}

\subsection{Proof of Proposition 1}
\label{sec:append_prop1}
We assume that the total transmit power is distributed equally among the jamming precoders for each user, such that, 
$	\sum_{n \in \mathcal{S}_{p,l}}||\mathbf{f}_{l,n}||^{2}=\frac{\bar{P}_{t}}{L}$, $\forall l \in \mathcal{L}$. Furthermore, the power allocated to the jamming precoders for AU-$l$ are divided equally among its pilot subcarriers, such that, $	||\mathbf{f}_{l,n}||^{2}=\frac{\bar{P}_{t}}{N_{p,l}}$,  $\forall n \in \mathcal{S}_{p,l}$.
The maximum jamming power on pilot subcarrier-$n$ of AU-$l$ is obtained as the solution of the problem
\begin{subequations}
	\begin{alignat}{3}
		\max_{\mathbf{p}}&     \quad  \mathbf{f}_{l,n}^{H}\mathbf{R}_{l,n}\mathbf{f}_{l,n} \label{eqn:thr_obj}\\
		\text{s.t.}& \quad  ||\mathbf{f}_{l,n}||_{2}^{2} \leq  \frac{\bar{P}_{t}}{N_{p,l}L} , \label{eqn:thr_constr}
	\end{alignat}
	\label{eqn:threshold}
\end{subequations}

Strong duality holds for problem as it has a quadratic objective and one quadratic inequality constraint and Slater’s condition holds \cite[Chapter 5]{boyd_2004}.
The necessary KKT conditions yield
\begin{subequations}
	\begin{alignat}{3}					
		&\mathbf{R}_{l,n}\mathbf{f}_{l,n}^{*}-\lambda^{*}\mathbf{f}_{l,n}^{*}=\mathbf{0}, \\ 
		&\lambda^{*}\left( ||\mathbf{f}_{l,n}^{*}||_{2}^{2}- \frac{\bar{P}_{t}}{N_{p,l}L}\right) =0, \label{eqn:problem_constr1_power}	 \\ 
		&\lambda^{*} \geq 0,  \\
		&\eqref{eqn:thr_constr}.
	\end{alignat}
\label{eqn:problem_constr1}	
\end{subequations}
\hspace{-0.1cm}The maximum is attained when the constraint \eqref{eqn:problem_constr1_power} is active and by the vector $\mathbf{f}_{l,n}^{*}$, which is in the same direction as the eigenvector of $\mathbf{R}_{l,n}$ that corresponds to the largest eigenvalue. Denoting the largest eigenvalue as $\sigma^{max}_{l,n}$, the solution of the problem \eqref{eqn:threshold} is obtained as \mbox{$ \frac{\bar{P}_{t}}{N_{p,l}L}\sigma^{max}_{l,n}$}. Therefore, we can conclude that the problem \eqref{eqn:problem1} is guaranteed to be feasible if $J^{thr}_{l,n} \leq \frac{\bar{P}_{t}}{N_{p,l}L}\sigma^{max}_{l,n}$, $\forall n \in \mathcal{S}_{p,l}$ and $\forall l \in \mathcal{L}$. We introduce the adjustment parameter $\rho \in \left[0,1 \right] $ to adjust the strictness of the threshold in the optimization problem, so that we write $J^{thr}_{l,n} = \rho\frac{\bar{P}_{t}}{N_{p,l}L}\sigma^{max}_{l,n}$.

One can note that the focused jamming power on an AU-$l$ may exceed the abovementioned value if $L>1$ and there exists another AU-$l^{\prime}$, with $l, l^{\prime} \in \mathcal{L}$ and $l\neq l^{\prime}$, for which the eigenvectors of  $\mathbf{R}_{l,n}$ and $\mathbf{R}_{l^{\prime},n}$ corresponding to $\sigma^{max}_{l,n}$ and $\sigma^{max}_{l^{\prime},n}$ are non-orthogonal. A simple but relevant example is a two-user scenario where $\mathbf{R}_{1,n}$ and $\mathbf{R}_{2,n}$ are identical. In such a scenario, the directions of the eigenvectors of $\mathbf{R}_{1,n}$ and $\mathbf{R}_{2,n}$ are identical, resulting in a maximum focused jamming power of $\frac{\bar{P}_{t}}{N_{p,l}}\sigma^{max}_{1,n}$ instead of $\frac{\bar{P}_{t}}{N_{p,l}2}\sigma^{max}_{1,n}$ for both AU-$1$ and AU-$2$. However, it is obvious that such occurrences depend on the characteristics of the AU channels and cannot be generalized in a way to guarantee the feasibility of the problem. Therefore, choosing $J^{thr}_{l,n} > \rho\frac{\bar{P}_{t}}{N_{p,l}L}\sigma^{max}_{l,n}$ does not guarantee a non-empty domain for problem \eqref{eqn:problem1}.

\subsection{Proof of Proposition 2}
\label{sec:append_prop2}
Let $\mathcal{N}\left(\wh{\mathbf{M}}^{H}\right)\backslash \mathbf{0}$ denote the null-space of $\wh{\mathbf{M}}^{H}$ excluding the all-zero vector. Note that $\mathcal{N}\left(\wh{\mathbf{M}}^{H}\right)$ is guaranteed to be non-empty for $N_{t}>N_{r}$. 
Choose a vector $\mathbf{f}^{*} \in \mathcal{N}\left(\wh{\mathbf{M}}^{H}\right)$ with $||\mathbf{f}^{*}||^{2}=\rho\frac{\bar{P}_{t}}{N_{p}L}<\frac{\bar{P}_{t}}{N_{p}L}$ for any $\rho \in [0,1)$. 
Recalling the definitions $\mathbf{S}^{*}=\mathbf{f}^{*}(\mathbf{f}^{*})^{H}$ and $\mathbf{\Phi}=\left(\hspace{-0.1cm}(1\minus\sigma_{pe}^{2}) \wh{\mathbf{M}}\wh{\mathbf{M}}^{H}\hspace{-0.1cm}+\hspace{-0.1cm}\sigma_{pe}^{2}N_{r,m}\mathbf{I}_{N_{t}}\hspace{-0.07cm}\right)$, one obtains
\begin{align}
	\mathrm{tr}\left(\mathbf{S}^{*}\mathbf{\Phi}\right)=(\mathbf{f}^{*})^{H}\mathbf{\Phi}\mathbf{f}^{*}=\rho\frac{\sigma_{pe}^{2}N_{r,m}\bar{P}_{t}}{N_{p}L}. \nonumber 
\end{align}

\subsection{Proof of Lemma 1}
\label{sec:append_lemma1}
It is well known that $\mathcal{N}\left(\wh{\mathbf{M}}^{H}\right)\backslash \mathbf{0}$ is spanned by the vectors $\mathbf{v}_{j}$ of the unitary matrix $\mathbf{V}$ for \mbox{$j \in \left\lbrace n_{m}+1, \ldots, N_{t} \right\rbrace $}. Then, for any \mbox{$\mathbf{f}^{*} \in  \mathcal{N}\left(\wh{\mathbf{M}}^{H}\right)\backslash \mathbf{0}$}, we can write 
\begin{align}
	\tilde{\mathbf{f}}^{*}=\mathbf{f}^{*}/||\mathbf{f}^{*}||=\mathbf{V}\mathbf{a},
	\label{eqn:p_nullM}
\end{align}
where $\mathbf{a} \in \mathbb{C}^{N_{t}}$ is the projection coefficient vector for $\tilde{\mathbf{f}}^{*}$ with elements $a_{i}=0$, \mbox{$\forall i \in \left\lbrace 1,2, \ldots,n_{m}\right\rbrace $}. We note that \mbox{$||a||^2=||\mathbf{V}^{H}\tilde{\mathbf{f}}^{*}||^{2}=||\tilde{\mathbf{f}}^{*}||^{2}=1$} since $\mathbf{V}$ is a unitary matrix. Substituting \eqref{eqn:p_nullM} into \eqref{eqn:rho_1} and using the eigen decomposition \mbox{$\mathbf{R}=\sum_{i}^{n_{g}}\sigma^{i} \mathbf{u}^{i}\left(\mathbf{u}^{i}\right)^{H} $}, we get
\begin{align}
	\rho &< \frac{1}{\sigma^{max}}\sum_{i}^{n_{g}}\sigma^{i}\hspace{-0.2cm} \sum_{j=n_{m}+1}^{N_{t}}\hspace{-0.2cm}a^{*}_{j} \left(\mathbf{v}^{j}\right)^{H} \mathbf{u}^{i}\left(\mathbf{u}^{i}\right)^{H}\hspace{-0.2cm}\sum_{j=n_{m}+1}^{N_{t}}\hspace{-0.2cm}a_{j} \mathbf{v}^{j}  \nonumber \\
	 &=\frac{1}{\sigma^{max}}\sum_{i=1}^{n_{g}}\sigma^{i}\hspace{-0.2cm}\sum_{j=n_{m}+1}^{N_{t}}\hspace{-0.2cm}||a_{j}\left( \mathbf{u}^{i}\right) ^{H} \mathbf{v}^{j}||^{2}. \nonumber
\end{align}

\subsection{Proof of Proposition 3}
\label{sec:append_prop3}
From the proof of Proposition 1, we know that the inequality $\mathbf{f}^{H}_{l,n}\mathbf{R}_{l,n}\mathbf{f}_{l,n}\geq\rho \frac{\bar{P}_{t}}{N_{p}L}\sigma^{max}_{l,n}$ can be satisfied with equality by \mbox{$\mathbf{f}^{\prime}_{l,n}=\sqrt{\rho \frac{\bar{P}_{t}}{N_{p}L}}\mathbf{u}^{max}_{l,n}$}. Then, the interference power at PU-$m$ due to $\mathbf{f}^{\prime}$ is given by
\begin{align}
	&\hspace{-0.35cm}\left( \mathbf{f}^{\prime}_{l,n}\right)^{H}\mathbf{\Phi}_{m,n} \mathbf{f}^{\prime}_{l,n} \nonumber \\
	&\hspace{-0.35cm}=\rho \frac{\bar{P}_{t}}{N_{p}L}\sum_{i=1}^{N_{t}}\lambda^{i}_{m,n}\left( \mathbf{u}^{max}_{l,n}\right)^{H}\mathbf{v}^{i}_{m,n}\left( \mathbf{v}^{i}_{m,n}\right)^{H}\mathbf{u}^{max}_{l,n}.
	\label{eqn:prop2_proof1}
\end{align}
The expression \eqref{eqn:prop2_proof1} shows that a solution for $\mathbf{f}$, which satisfies both jamming and interference power constraints, exists for the general case if $I^{thr}_{l,m,n}$ is chosen according to \eqref{eqn:prop2_proof1}.

It is shown in Section~\ref{sec:feasibility_analysis} that for the case with Lagrangian multipliers $\alpha^{*}>0$, $\beta^{*}=\gamma^{*}=0$, the jamming power threshold can be set as $I^{thr}_{l,m,n}=0$ if $\sigma_{pe}=0$ and \eqref{eqn:rho_2} holds for any $\mathbf{a} \in \mathbb{C}^{N_{t}}$ with elements $a_{i}=0$, \mbox{$\forall i \in \left\lbrace 1,2, \ldots,n_{m}\right\rbrace $}. We define the set of vectors \mbox{$\mathcal{E}=\left\lbrace \mathbf{e}_{n_{m}+1}, \mathbf{e}_{n_{m}+2}, \ldots, \mathbf{e}_{N_{t}}\right\rbrace $}. It is clear that $\mathcal{E}$ is a subset of the set of all possible values of $\mathbf{a}$, thus setting $I^{thr}_{l,m,n}=\mu$ for any $\mu \in \mathbb{R}_{0+}$ is feasible if the condition \eqref{eqn:rho_2} is satisfied for any $\mathbf{a} \in \mathcal{E}$. 

From the arguments above and $I^{thr}_{l,m,n}\geq0$, $\forall n \in \mathcal{S}_{p,l}$ $\forall l \in \mathcal{L}$, $\forall m \in \mathcal{M}$ it is straightforward to show that $I^{thr}_{m,n}=\sum_{l\in \mathcal{L}}I^{thr}_{l,m,n}$, where $I^{thr}_{l,m,n}=\psi(\rho, \bar{P}_{t}, N_{p}, L, \mathbf{R}_{l,n}, \mathbf{\Phi}_{m,n},\mu,\sigma_{pe})$ guarantees a non-empty domain for problem \eqref{eqn:problem2_final_nonconvex}. 
  
\subsection{Proof of Corollary 1}
\label{sec:append_cor1}
We consider the projection of $\mathbf{u}^{max}$ on the vector space spanned by the columns of $\mathbf{V}$, so that
\begin{align}
	\mathbf{u}^{max}&=\mathbf{V}\mathbf{b}=\sum_{i=1}^{N_{t}}b_{i}\mathbf{v}^{i},
	\label{eqn:umax_nullM}
\end{align} 
with $||\mathbf{b}||^{2}=1$ as demonstrated in the proof of Lemma~1. We substitute \eqref{eqn:umax_nullM} into the $I^{thr}_{l,m,n}$ expression in Alg.~\ref{alg:algorithm_thr} to obtain
\begin{align}
	I^{thr}_{l,m,n}&=\rho \frac{\bar{P}_{t}}{N_{p}L}\sum_{i=1}^{n_{m}}\lambda^{i} \sum_{j=1}^{N_{t}}b_{j}^{*}\left(\mathbf{v}^{j}\right)^{H}\mathbf{v}^{i}\left(\mathbf{v}^{i}\right)^{H}\sum_{j=1}^{N_{t}}b_{j}\mathbf{v}^{j} \nonumber\\
	&=\rho \frac{\bar{P}_{t}}{N_{p}L}\sum_{i=1}^{n_{m}}\lambda^{i}|b_{i}|^{2} \leq \rho \frac{\bar{P}_{t}}{N_{p}L}\lambda^{max}\sum_{i=1}^{n_{m}}|b_{i}|^{2}\nonumber\\
	&\leq \frac{\bar{P}_{t}}{N_{p}L}\lambda^{max}, \nonumber
\end{align}
where the last inequality is due to $\rho\leq 1$ and \mbox{$\sum_{i=1}^{n_{m}}|b_{i}|^{2}\leq\sum_{i=1}^{N_{t}}|b_{i}|^{2}=1$}.
For the case the condition \eqref{eqn:rho_2} is satisfied, the proof is straightforward for any $\mu\in \mathbb{R}_{0+}$ and $\mu \leq \frac{\bar{P}_{t}}{N_{p}L}\lambda^{max}$. 

\subsection{Proof of Corollary 2}
\label{sec:append_cor2}
We drop the subcarrier and user indexes for simplicity.
First, consider the problem \eqref{eqn:problem_analysis_sdp} with $\alpha^{*}>0$, $\gamma^{*}=0$, $\beta^{*}>0$ satisfying \mbox{$\mathbf{a}^{H}\left( \mathbf{\Phi}-\beta^{*}\mathbf{R}_{g}\right)\mathbf{a} = 0$} for any $\mathbf{a} \in \mathbb{C}^{N_{t}}$. Then,

\begin{align}
	I^{*}&=\rho \frac{\bar{P}_{t}}{N_{p}L}\left( \mathbf{u}^{max}\right)^{H}\mathbf{\Phi}\mathbf{u}^{max} \nonumber \\
	&=\beta^{*} \rho \frac{\bar{P}_{t}}{N_{p}L}\left( \mathbf{u}^{max}\right)^{H}\mathbf{R}\mathbf{u}^{max} \nonumber \\
	&=\beta^{*} \rho \frac{\bar{P}_{t}}{N_{p}L}\sigma^{max},
\end{align} 
which is equal to the threshold in expression \eqref{eqn:thr1}.

The proof for the case with $\alpha^{*}>0$, $\beta^{*}=\gamma^{*}=0$ (which yields \eqref{eqn:thr2}) is straightforward by setting $\mu=0$, and thus omitted.

\end{document}